\documentclass[12pt]{article}

\usepackage[unicode=true,
 bookmarks=false,
breaklinks=false,pdfborder={0 0 1},backref=false,colorlinks=false]
 {hyperref}
\usepackage[a4paper, total={210mm,297mm},margin=2.0cm]{geometry}

\usepackage{cite}

\usepackage{amssymb}
\usepackage{amsmath}
\usepackage{pifont}
\usepackage{graphicx}
\usepackage{epstopdf}
\usepackage{hyperref}
\usepackage{color}
\usepackage{gensymb}
\usepackage{subfig}
\usepackage{times}
\usepackage{rotating}
\usepackage{lscape}
\usepackage{tabularx}
\usepackage{multirow}
\usepackage[normalem]{ulem}
\usepackage{caption}
\usepackage{tabularx}
\usepackage{multirow}
\usepackage{epstopdf}
\usepackage[utf8]{inputenc}
\usepackage{nomencl}
\usepackage[all]{xy}
\usepackage{tikz}
\usetikzlibrary{arrows,backgrounds,calc,graphs,shapes}
\usepackage{chemfig}
\usepackage{array}

\newcommand{\ba}{\begin{array}}
\newcommand{\ea}{\end{array}}
\newcommand{\be}{\begin{equation}}
\newcommand{\ee}{\end{equation}}
\newcommand{\bc}{\begin{center}}
\newcommand{\ec}{\end{center}}
\newcommand{\bdm}{\begin{displaymath}}
\newcommand{\edm}{\end{displaymath}}

\providecommand{\keywords}[1]{\textit{Keywords:} #1}

\begin{document}

\title{Modelling Phase Separation In Amorphous Solid Dispersions}

\author{Martin Meere$^{a,}$\footnote{Corresponding author. {\em Email:} martin.meere@nuigalway.ie} , 
            Giuseppe Pontrelli$^{b}$, Sean McGinty$^{c}$ \\  \\  \\
            $^{a}$School of Mathematics, NUI Galway, University Road, \\
            Galway, Ireland \\
            $^{b}$ Istituto per le Applicazioni del Calcolo, CNR, Rome, Italy \\
            $^{c}$ Division of Biomedical Engineering, University of Glasgow, \\
            Glasgow, G12 8QQ, UK}

\vspace{0.3cm}

\date{}

\maketitle

\begin{abstract} 
Much work has been devoted to analysing thermodynamic models for solid dispersions 
with a view to identifying regions in the phase diagram where amorphous phase separation 
or drug recrystallization can occur. However, detailed partial differential equation 
non-equilibrium models that track the evolution of solid dispersions in time and space 
are lacking. Hence theoretical predictions for the timescale over which phase separation 
occurs in a solid dispersion are not available. In this paper, we address some of these 
deficiencies by (i) constructing a general multicomponent diffusion model for a dissolving 
solid dispersion; (ii) specializing the model to a binary drug/polymer system in storage; 
(iii) deriving an effective concentration dependent drug diffusion coefficient for the binary 
system, thereby obtaining a theoretical prediction for the timescale over which phase 
separation occurs; (iv) calculating the phase diagram for the Felodipine/HPMCAS 
system; and (iv) presenting a detailed numerical investigation of the Felodipine/HPMCAS system 
assuming a Flory-Huggins activity coefficient. The numerical simulations exhibit numerous 
interesting phenomena, such as the formation of polymer droplets and strings, Ostwald 
ripening/coarsening, phase inversion, and droplet-to-string transitions. A numerical
simulation of the fabrication process for a solid dispersion in a hot melt extruder was also 
presented.
\end{abstract}

\noindent \keywords{amorphous solid dispersion, phase separation, mathematical model,
                 drug diffusion}

\section{Introduction}

Drugs that are delivered orally via a tablet should ideally be readily soluble in water. Drugs 
that are poorly water-soluble tend to pass through the gastrointestinal tract before they 
can fully dissolve, and this typically leads to poor bioavailability of the drug. Unfortunately,
many drugs currently on the market or in development are poorly water-soluble, and this 
presents a serious challenge to the pharmaceutical industry. Many strategies have been 
developed to improve the solubility of drugs, such as the use of surfactants, cocrystals, 
lipid-based formulations, and particle size reduction. The literature on this topic is extensive, 
and recent reviews can be found in \cite{Savjani:2012,Williams:2013,Kalepu:2015}.      
  
\begin{figure}[ht] 

\begin{center}
\includegraphics[width=10cm,height=8cm]{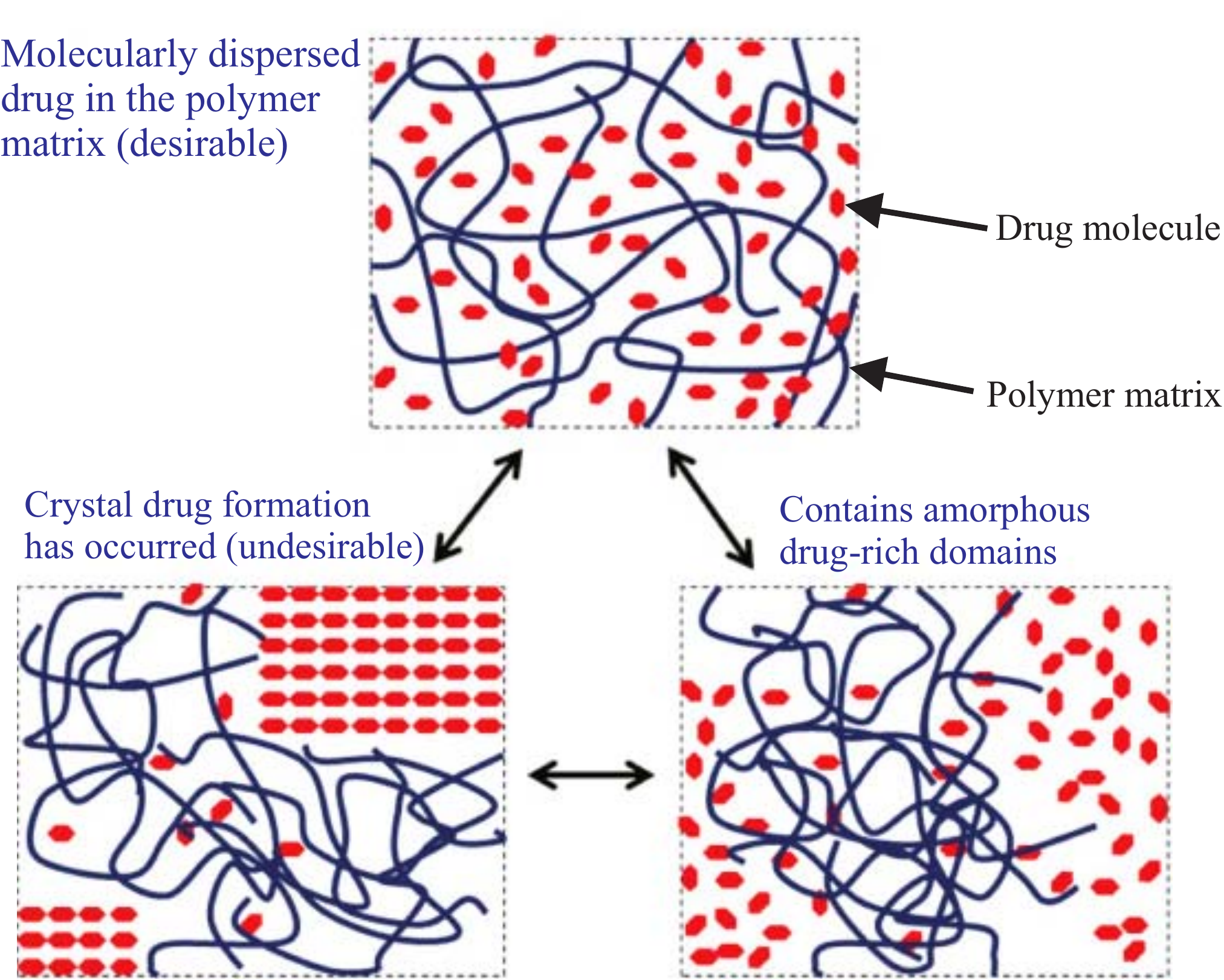}
\end{center}
\caption{Adapted from \cite{Huang:2014}. In this figure, we show three possible structures
for a polymer/drug dispersion. Top: Here the drug is in the molecularly dispersed state, which
is usually desirable for a solid dispersion. Bottom left: Here the dispersion contains drug in the 
crystalline form. Bottom right: Here the dispersion contains amorphous drug-rich domains.} 
\label{f1} 
\end{figure}

One particularly effective strategy to improve drug solubility is to use a {\em solid dispersion}
\cite{Brough:2013,Huang:2014,Dhirendra:2009}. A solid dispersion typically consists of a 
hydrophobic drug embedded in a hydrophilic polymer \cite{Souery:2018,Garcia:2019} matrix, where 
the matrix can be either in the amorphous or crystalline state. The drug is preferably in a molecularly 
dispersed state, but may also be present in amorphous particles or even in the crystalline form (though 
this is usually undesirable); see Figure \ref{f1}. The drug release concept for most solid dispersions 
is based on the so-called {\em spring and parachute effect} \cite{Brouwers:2009}. When the 
drug and the hydrophilic polymer dissolve in solution, a supersaturated drug solution is quickly 
created (the spring). Although the drug concentration then subsequently decreases, the rate 
of decrease is slowed by drug-polymer interactions in the dispersion, so that the drug can be 
present at supersaturated levels in the solution for a period of some hours (the parachute). This 
results in improved bioavailability of the drug when the solid dispersion dosage form is taken 
orally. 

Drug loading in most dispersions greatly exceeds the equilibrium solubility in the polymer 
matrix for typical storage temperatures. Hence these systems are usually unstable, with 
phase separation eventually occurring \cite{Huang:2014}. In such cases, the drug will 
eventually crystallise out or form an amorphous phase separation. However, if the dispersion 
is stored well below the glass transition temperature  \cite{Doi:1996} for the polymer, and 
is kept dry, this can happen extremely slowly. The system is then for all practical purposes 
stable, and is said to be metastable. The humidity of the storage environment can be an 
issue because even small amounts of moisture can significantly affect the glass transition 
temperature. Hence polymers that have high glass transition temperatures and that are 
resistant to  water absorption have become popular. An example of one such polymer is 
Hydroxypropyl Methylcellulose Acetate Succinate (HPMCAS).  

Phase separation of solid dispersions in storage is clearly undesirable from the point of 
manufacturers. Hence much work has been devoted to constructing phase diagrams for 
solid dispersions with a view to identifying regimes where drug recrystallization or 
amorphous phase separation can occur. These phase diagrams are constructed with the
aid of thermodynamic models. The most widely used thermodynamic model in this context 
is the Flory-Huggins model \cite{Flory:1942,Huggins:1941,Hiemenz:2007} for polymer 
solutions. 

Flory-Huggins theory is a lattice-based model in which the drug and polymer are confined
to live on a regular lattice. Flory-Huggins theory is an extension of regular solution 
theory, as explained in Chapter 7 of \cite{Hiemenz:2007}. In the context of a drug/polymer 
system, each drug molecule is taken to occupy one lattice site and each polymer segment 
is taken to occupy $m\gg 1$ sites. Under a number of further simplifying assumptions
\cite{Hiemenz:2007}, the change in entropy and enthalpy associated  with the mixing of 
the polymer and drug are calculated. With these in hand, the change in bulk Gibbs free 
energy ($\Delta g_{b}^{mix}$) per mole associated with mixing is readily calculated, and is 
found to be
\begin{equation}
\frac{\Delta g_{b}^{mix}}{RT} = X_{d}\ln(\phi_{d}) + X_{p}\ln(\phi_{p}) 
 + \chi_{dp} X_{d}\phi_{p},    \label{eq:deltagmix}
\end{equation}  
where $R$ is the gas constant, $T$ is the temperature, $X_{d},X_{p}=1-X_{d}$ are the mole fractions 
of the drug and polymer, respectively, and $\phi_{d},\phi_{p}$ are the volume fractions 
of the drug and polymer, respectively. The quantity $\chi_{dp}$ is referred to as the Flory-Huggins
interaction parameter, and it is discussed further below. The mole fractions and volume fractions 
are related via the formulae
\begin{equation} 
\phi_d = \frac{X_d}{X_d+m X_p}, \hspace{0.35cm}\phi_p = \frac{m X_p}{X_d+m X_p}.
\label{eq:volumefractions}
\end{equation}
When the model is applied to real binary systems, $m$ can be calculated using the formula
\begin{equation}
m = \frac{V_{p}}{V_{d}}       \label{eq:volumeratio} 
\end{equation}   
where $V_{p}$, $V_{d}$ (molar$^{-1}$) are the molar volumes of the polymer and drug, 
respectively.  

The mixing of the polymer and drug is spontaneous if $\Delta G_{mix}<0$. 
The Flory-Huggins parameter $\chi_{dp}$ takes the form  
\begin{equation}
  \chi_{dp} = \rho^{2} \left( w_{dp} - \frac{w_{dd}+w_{pp}}{2} \right)  
  \label{eq:chienergies}  
\end{equation}   
where $\rho^{2}$ is a positive parameter, and $w_{dp}$, $w_{dd}$, $w_{pp}$ give 
measures of the drug-polymer, drug-drug and polymer-polymer interaction energy, 
respectively. If $\chi_{dp}<0$ then $w_{dp}<(w_{dd}+w_{pp})/2$ indicating that
that the mixed state has lower energy than the separated pure states, so that mixing
is favoured. Conversely, $\chi_{dp}>0$ is indicative of demixing being favoured. However,
these statements are indicative rather than precise, as will be explained in Section 3. 
We should also note that $\chi_{dp}$ is temperature dependent, and is usually given 
the empirical form
\begin{equation}
\chi_{dp}(T) = \frac{\alpha}{T} + \beta    
\label{eq:xhidpT} 
\end{equation}
where $\alpha,\beta$ are constants. 

Flory-Huggins theory has frequently been used to analyse the stability of binary 
solid dispersion systems in storage; see, for example, \cite{Djuris:2013,Tian:2013, Tian2:2013,
Bansal:2016,Chan:2015,Altamimi:2016,Chakravarty:2017,Lin:2010,Wlodarski:2015,
Zhao:2011}. In many of these studies, the Flory-Huggins interaction parameter is 
first estimated using the melting point depression method \cite{Marsac:2006}, or 
using the Hildebrand and Scott method \cite{Hildebrand:1964}, which involves the
estimation of solubility parameters. Once estimates for $\chi_{dp}(T)$ have been 
obtained, the Gibbs free energy of mixing $\Delta G_{mix}$ can be calculated, 
which in turn enables the construction of phase diagrams for the systems. Phase 
diagrams assist with the identification of regions in composition-temperature space 
where the system is prone to recrystallization or amorphous phase separation.

The models we shall develop in the current study are generic and are not 
tied to making a specific choice of statistical model. However, given the particular 
importance of Flory-Huggins theory in applications, we shall derive detailed results 
for this case. Also, all of our numerical illustrations are calculated within the 
context of Flory-Huggins theory. It should be emphasized that Flory-Huggins theory 
does involve quite a number of simplifying assumptions which are not appropriate 
for some systems; see \cite{Anderson:2018} for a recent critique of the model. 

\section{Theoretical formulation}  
  
\subsection{A multicomponent diffusion model for solid dispersions} 

We develop a multicomponent diffusion model for the evolution of the concentrations of 
the components constituting a solid dispersion. We suppose for the moment that there 
are $p$ components. However, in the analysis we shall consider in the current study, 
we will in fact have $p=2$, with one of the components being the polymer, and the other 
being the drug. For a dissolving solid dispersion, there are three components $p=3$: the 
polymer, the drug, and the solvent. 

The chemical potential $\mu_i$ (J/mole) of species $i$ $(i=1,2,...,p)$ gives the 
Gibbs free energy per mole of species $i$, and is given here by 
(\cite{Smith:2014})
\begin{equation}
\mu_{i} = \mu_{i}^{b}  - \epsilon_{i}^{2} \nabla^{2} X_{i}  \label{eq:cp}
\end{equation}
where  
\begin{equation}
\mu_{i}^{b} = \mu_{i}^{0} + RT\ln (a_{i})    \label{eq:muib}
\end{equation}
and where $\mu_{i}^{b}$ is the bulk chemical potential of species $i$, $\mu_{i0}$ 
is the chemical potential of species $i$ in the pure state, $a_{i}$ is the activity of species 
$i$, and the term involving $\epsilon_{i}^{2}>0$ (m$^{2}$J/mole) penalises the formation 
of phase boundaries (\cite{Binder:2001}, \cite{Saylor:2007},\cite{Saylor:2011},
\cite{Zhu:2016}, \cite{Zhu:2017}). The parameters $\epsilon_{i}^{2}$ are referred to as gradient 
energy coefficients (\cite{Cahn:1958}, \cite{Provatas:2010}). Here $X_i$ is the molar fraction of 
species $i$ ($i=1,2,...,p$), and the activities can depend on these molar fractions, so that 
\[ a_i = a_i(X_1,X_2,...,X_p). \]
The molar fraction is related to the molar concentration via
\begin{equation} 
X_i = V_i c_i   \label{eq:mvolume}  
\end{equation}  
where $V_i$ (molar$^{-1}$) is the molar volume of species $i$. The flux of species $i$ 
(molar$\cdot$m/s)  is given by
\begin{equation} 
{\bf J}_{i} = c_{i} {\bf v}_{i}       \label{eq:fluxi} 
\end{equation}
where $c_i$ (molar), ${\bf v}_i$ (m/s) give the molar concentration and drift 
velocity, respectively, of species $i$. The drift velocity ${\bf v}_{i}$ gives the average 
velocity a particle of species $i$ attains due to the diffusion force acting on it, 
and is given here by
\begin{equation} 
{\bf v}_i = M_i{\bf {\cal F}}_i = -M_i \nabla \mu_i 
\label{eq:drifti}
\end{equation}
where $M_i$ (mole$\cdot$s/kg), ${\bf {\cal F}}_i$ (J/[m$\cdot$mole]) give the 
mobility and diffusion force, respectively, for species $i$. Equations (\ref{eq:fluxi})
and  (\ref{eq:drifti}) give
\begin{equation}
 {\bf J}_i = -M_i c_i \nabla\mu_i.  \label{eq:Mici}
\end{equation}   
Conservation of mass for species $i$ implies that 
\begin{equation}
 \frac{\partial c_i}{\partial t} +  \nabla\cdot {\bf J}_i = 0 \label{eq:comii}
\end{equation}
and using (\ref{eq:Mici}) now gives 
\begin{equation*} 
\frac{\partial c_i}{\partial t} = \nabla \cdot \left( M_{i}c_{i}\nabla\mu_{i} \right) 
\end{equation*}
or equivalently 
\begin{equation} 
\frac{\partial X_i}{\partial t} = \nabla \cdot \left( D_{i}X_{i}\nabla
 \left\{\frac{\mu_{i}-\mu_{i}^{0}}{RT}\right\} \right) 
\label{eq:numform}
\end{equation}
with 
\begin{equation}
\frac{\mu_{i} - \mu_{i}^{0}}{RT} = \ln (a_{i}) - \delta_{i}^{2} \nabla^{2} X_{i}  
\label{eq:cpboxed}
\end{equation}
for $i=1,2,...,p$, and where $\delta_{i}^{2}=\epsilon_{i}^{2}/RT>0$ (m$^{2}$/molar),
and 
\[ D_i = M_i RT \hspace{1.0cm}\mbox{(Einstein relation)} \] 
is the self-diffusion coefficient for species $i$.

The model formulation given by (\ref{eq:numform}) and (\ref{eq:cpboxed}) based on 
chemical potentials will be used for the numerical scheme described in Section \ref{numericss}. 
However, it is also of value to develop a formulation involving diffusion coefficients since these
yield immediate information regarding timescales for transport processes, and will also the enable
the development of analytical results via a linearization process. 

\vspace{0.4cm}

\noindent {\em Diffusion Coefficients} 

\vspace{0.4cm}

Using (\ref{eq:cp}), (\ref{eq:muib}) and (\ref{eq:Mici}) gives
\[ {\bf J}_i = -M_i c_i \nabla\mu_i  
       = -M_i c_i \left(\frac{RT}{a_i} \nabla a_i 
            -  \epsilon_{i}^{2}\nabla (\nabla^{2}X_{i})\right) \]
and then using the fact that the activities depend on the molar fractions gives 
\begin{equation}
 {\bf J}_i = -M_i c_i \left(\frac{RT}{a_i} \sum_{j=1}^{p}\frac{\partial a_i}{\partial X_{j}}
               \nabla X_{j} - \epsilon_{i}^{2}\nabla(\nabla^{2} X_{i}) \right).
\label{eq:fluxii}
\end{equation}
 Using (\ref{eq:mvolume}),
we can now write (\ref{eq:fluxii}) as
\begin{equation}
 {\bf J}_i = - \sum_{j=1}^{p} D_{ij}\nabla c_j + 
          D_{i}\varepsilon_{i}^{2}c_{i} \nabla (\nabla^{2}c_{i}) 
\label{eq:fluxiii}
\end{equation}
where $\varepsilon_{i}^{2}=V_{i}\delta_{i}^{2}$ and where the diffusion coefficients
$D_{ij}$ (m$^{2}$/s) are given by
\begin{equation}
D_{ij} = D_i \frac{V_j}{V_i}\frac{X_i}{a_i}\frac{\partial a_i}{\partial X_j}.  
 \hspace{1.0cm} i,j=1,2,...,p   \label{eq:diffcoeffs}  
\end{equation}
Conservation of mass (\ref{eq:comii}) then implies that (reverting to molar 
fractions)
\begin{equation} 
\frac{\partial X_i}{\partial t} = \nabla\cdot
\left(\sum_{j=1}^{p} \frac{V_i}{V_j} D_{ij}({\bf X})  \nabla X_j 
 - D_{i}\delta_{i}^{2} X_{i} \nabla (\nabla^{2} X_i) \right) 
\hspace{1.0cm} i=1,2,...,p \label{eq:mcdm}
\end{equation}
where ${\bf X}=(X_1,X_2,...,X_p)$, and where we have included the concentration 
dependence of the diffusion coefficients $D_{ij}$ here to emphasise that this system 
is in general a coupled system of nonlinear diffusion equations. It should benoted that 
the equations (\ref{eq:mcdm}) are not independent since $\sum_{i=1}^{p}X_{i}=1$, 
and so it is sufficient to solve for $p-1$ concentrations only. 

\subsection{Activity coefficients} The activities $a_i$ are usually 
written as  
\[  a_i = \gamma_i X_i \]
where the $\gamma_i=\gamma_i(X_1,X_2,...,X_p)$ are referred to as {\em activity
coefficients}. Equations (\ref{eq:diffcoeffs}) now give
\begin{equation}
 D_{ij} = D_{i}\frac{V_j}{V_i}\left(\delta_{ij} + \frac{X_i}{\gamma_i}
  \frac{\partial\gamma_i}{\partial X_j} \right) \hspace{1.0cm}  i,j=1,2,...,p  
  \label{eq:diffg}  
\end{equation} 
where $\delta_{ij}$ is the Kronecker delta. 

The details of the interactions between the species in solution are 
captured in the modelling by choosing appropriate forms for the activity
coefficients $\gamma_i=\gamma_i(X_1,X_2,...,X_p)$. The construction of 
appropriate forms for the $\gamma_i$ for various solutions is a large
subject with a large literature; see, for example, the books 
\cite{Georgis:2010} and \cite{Prausnitz:1999}.

\subsection{The storage problem for a binary mixture} 

In the current study, we shall be modelling the behaviour of solid dispersions in
storage. In this case, we have $p=2$, with the label 1 referring to the drug and
the label 2 referring to the polymer. However, for transparency, we choose here 
to use the labels $d,p$ rather than $1,2$, where $d$ stands for drug, and 
$p$ for polymer. Then using (\ref{eq:mcdm}) and the fact that $X_p = 1 - X_d$, 
we have 
\begin{equation}
\frac{\partial X_d}{\partial t} = \nabla \cdot \left\{ D_{\mbox{\scriptsize eff}}(X_d)
\nabla X_d - D_{d}\delta_{d}^{2}X_d  \nabla \left(\nabla^{2}X_d\right) \right\}. 
\label{eq:D11D12}
\end{equation}
where the effective concentration-dependent diffusion coefficient for the drug in the
solid dispersion is  
\begin{equation}
D_{\mbox{\scriptsize eff}}(X_d) = D_{dd}(X_d)-V_d D_{dp}(X_d)/V_p 
 = D_{d}\left\{1 + \frac{X_{d}}{\gamma_{d}}
\left[ \frac{\partial \gamma_d}{\partial X_d} -\frac{\partial \gamma_d}{\partial X_p}  
\right] \right\}.
\label{eq:effdiffusivity}
\end{equation} 
For the particular case of a binary Flory-Huggins theory (see Section 1), the 
activity coefficients are given by
\begin{eqnarray}
&&  \ln(\gamma_{d})  = \ln\left(\frac{\phi_d}{X_d}\right) 
+1-\frac{\phi_d}{X_d} + \chi_{dp} \phi_{p}^{2}, \label{eq:gum1} \\
&&  \ln(\gamma_{p})  = \ln\left(\frac{\phi_p}{X_p}\right) 
+1-\frac{\phi_p}{X_p} + m\chi_{dp} \phi_{d}^{2}, \label{eq:gum2}
\end{eqnarray}
where the volume fractions $\phi_{d},\phi_{p}$ are given by (\ref{eq:volumefractions}). 
Substituting (\ref{eq:gum1}) in (\ref{eq:effdiffusivity}) gives 
\begin{equation}
D_{\mbox{\scriptsize eff}}(X_d) = D_d \left\{ {(m-(m-1)X_d)(m^2 - (m^{2} - 1)X_d) 
                           - 2 \chi_{dp} m^{2} X_d (1 - X_d) 
                             \over (m - (m - 1) X_d)^3} \right\}.   \label{eq:effdiffN}   
\end{equation}
It is more instructive to write this expression in terms of the volume fraction of drug. 
Writing $D_{\mbox{\scriptsize eff}}(X_d)=\tilde{D}_{\mbox{\scriptsize eff}}(\phi_d)$,
we obtain 
\begin{equation}
\boxed{\tilde{D}_{\mbox{\scriptsize eff}}(\phi_d)=D_d (1 + (m-1)\phi_{d})
  \left( 1 + \left(\frac{1}{m} - 1\right)\phi_d  - 
 2\chi_{dp}\phi_{d}(1 - \phi_{d})  \right)} .   \label{eq:effdiffNphi}   
\end{equation}
This expression is particularly useful because it yields insight into how the mobility
of the drug in the dispersion depends on the length of the polymer chains ($m$), the 
dispersion composition ($\phi_d$), and the character of the drug-polymer interaction 
($\chi_{dp}$). We shall analyze this expression further in Section 3, and also show 
how it can be used to calculate the timescale over which phase separation may occur. 

An equivalent formulation for the Flory-Huggins model involving the chemical potential 
for the drug $\mu_d$ is given by (see (\ref{eq:numform}) and (\ref{eq:cpboxed}) above): 
\begin{equation} 
\frac{\partial X_d}{\partial t} = \nabla \cdot \left( D_{d}X_{d}\nabla\psi\right) \label{eq:eqNd}
\end{equation}
where 
\begin{equation}
  \psi = \frac{\mu_{d}-\mu_{d0}}{RT}  \label{eq:defpsi}  
\end{equation}  
and with 
\begin{equation}
\psi = \ln\left(\frac{X_{d}}{m - (m - 1)X_{d}} \right)
+ \frac{(m - 1)(1 - X_{d})}{m - (m - 1)X_{d}} + 
\chi_{dp}m^{2}\left(\frac{1 - X_{d}}{m - (m - 1)X_{d}}\right)^{2} - 
\delta_{d}^{2} \nabla^{2} X_{d}.  \label{eq:eqchi}
\end{equation}

We suppose that the solid dispersion occupies a two-dimensional region $\Omega$. 
The governing equation for the drug concentration in $\Omega$ may be written in 
the conservation form  
\[ \frac{\partial X_d}{\partial t} + \nabla\cdot {\bf J}_{d} = 0, \] 
where the drug flux ${\bf J}_{d}$ is given by 
\begin{equation}
  {\bf J}_{d} = - D_{\mbox{\scriptsize eff}}(X_d) \nabla X_{d} + 
 D_{d}\delta_{d}^{2}X_d \nabla (\nabla^2 X_d).   \label{eq:fluxd}
\end{equation} 
We need to supplement the governing equation in $\Omega$ with boundary conditions 
on $\partial\Omega$, and we choose these here to be
\begin{equation}
  {\bf J}_{d}\cdot {\bf n} = 0,  \hspace{0.5cm} 
   \nabla X_{d}\cdot {\bf n}  = 0  \hspace{0.5cm}
   \mbox{ on }\hspace{0.5cm}\partial\Omega .   \label{eq:bcs2d}   
\end{equation}
The first of these conditions ${\bf J}_{d}\cdot {\bf n}=0$ implies that the drug cannot 
penetrate the boundary of the domain. The other condition $\nabla X_{d}\cdot {\bf n}=0$ 
is the natural boundary condition for the variational formulation of the problem, and it implies 
that the interfaces between the polymer rich and the drug rich domains meet the boundary 
at right angles.

Finally, to obtain a well-posed problem, we need to impose an initial condition and we choose 
this here to take the form 
\begin{equation}
   X_{d}(x,y,t=0) = X_{d}^{0}(x,y) \hspace{0.5cm}\mbox{ for }\hspace{0.5cm}
  (x,y)\in\Omega ,    \label{eq:icform} 
\end{equation}   
where  $X_{d}^{0}(x,y)$ is a given function. 

Gathering together the governing equation, the boundary conditions and the initial condition, 
we obtain the following initial boundary value problem: 
\begin{eqnarray}
&& \frac{\partial X_d}{\partial t} = \nabla\cdot\left\{ D_{\mbox{\scriptsize eff}}(X_d)\nabla X_d 
 - D_{d}\delta_{d}^{2}X_d \nabla\left(\nabla^{2}X_d \right) \right\} \hspace{0.5cm}
 \mbox{ in }\hspace{0.5cm} \nonumber   \Omega, \\
&& \nabla X_{d}\cdot {\bf n} = 0,  \hspace{0.5cm} 
   \nabla (\nabla^{2} X_{d})\cdot {\bf n} = 0  \hspace{0.5cm}
   \mbox{ on }\hspace{0.5cm}\partial\Omega , \label{eq:2divp} \\
&& X_{d}(x,y,t=0) = X_{d}^{0}(x,y) \hspace{0.5cm}\mbox{ for }\hspace{0.5cm}
  (x,y)\in\Omega . \nonumber 
\end{eqnarray}

\subsection{Phase separation in a Flory-Huggins binary mixture \label{sec:phaseseparation}} 

\noindent {\em The bulk free energy and spinodal decomposition}

\vspace{0.2cm} 

Spinodal decomposition for binary systems has been long understood using thermodynamic 
reasoning, and is well described elsewhere; see, for example,  Chapter 5 of \cite{Porter:2009}, 
Chapter 7 of \cite{Hiemenz:2007}, or \cite{Elbert:2011}. Hence our description of the background 
theory here will be quite brief, and we will emphasise instead the particular details for the 
Flory-Huggins system.  

The bulk free energy density $g_b$ for the binary mixture constituting the solid 
dispersion is given by (\cite{Atkins:2009})
\begin{equation}
g_b = \mu_d^b X_d + \mu_p^b X_p, \label{eq:freebulk}
\end{equation}  
where  $\mu_d^b,\mu_p^b$ give the bulk chemical potential of the drug and polymer,
respectively, and where
\[  \mu_i^b = \mu_i^0 + RT \ln(a_i) \hspace{0.8cm}\mbox{ for }\hspace{0.4cm}i=d,p.   \]
This leads to  
\[ g_b = \mu_d^0 X_d + \mu_p^0 X_p +RT( X_d\ln(X_d)+X_p\ln(X_p) ) 
            + RT( X_d\ln(\gamma_d)+X_p\ln(\gamma_p) ). \]
The Gibbs free energy of mixing $\Delta g_b^{mix}$ is given by
\[      \Delta g_b^{mix} =    g_b - \mu_d^0 X_d - \mu_p^0 X_p  = RT( X_d\ln(X_d)+X_p\ln(X_p) ) 
            + RT( X_d\ln(\gamma_d)+X_p\ln(\gamma_p) ). \]
If we now use (\ref{eq:gum1}) and (\ref{eq:gum2}) and the fact that $X_p=1-X_d$, we
arrive at 
\begin{equation}
\frac{\Delta g_b^{mix}}{RT} =  X_d\ln\left(\frac{X_d}{m - (m - 1)X_d}\right)  
    + (1 - X_d) \ln\left(\frac{m(1 - X_d)}{m - (m - 1)X_d}\right) + 
       \frac{\chi_{dp}m X_d(1 - X_d)}{m - (m - 1)X_d} . 
\label{eq:Gb}
\end{equation}   
\begin{figure}[ht] 
\begin{center}
\includegraphics[width=15.4cm,height=6.3cm]{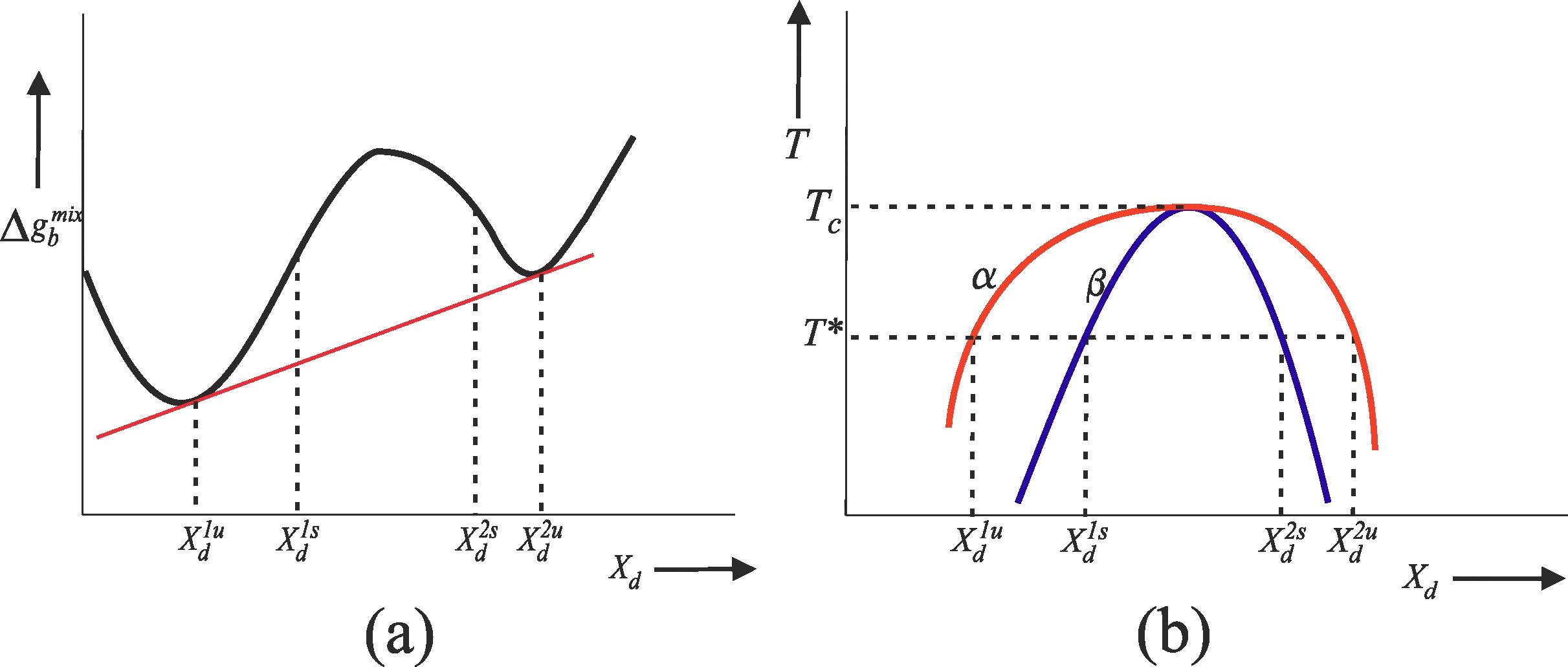}
\end{center}
\caption{(a) Plot of the bulk free energy of mixing $\Delta g_b^{mix}$ as a function of 
a drug molar fraction $X_d$. The spinodal points $X_{d}^{1s}$, $X_{d}^{2s}$ are the 
solutions to $d^{2}(\Delta g_{b}^{mix})/dX_{d}^{2}=0$. In the spinodal region 
$(X_{d}^{1s},X_{d}^{2s})$, we have $d^{2}(\Delta g_{b}^{mix})/dX_{d}^{2}<0$ 
and $D_{\mbox{\scriptsize eff}}(X_d)<0$. 
(b) Phase diagram for the binary mixture. Here $\alpha$ is the coexistence curve, 
$\beta$ is the spinodal curve, $T^{*}$ is the temperature for the free energy density 
diagram in (a), and $T_{c}$ is the critical temperature above which the dispersion is 
homogeneous.} 
\label{fspin} 
\end{figure}  
In Figure \ref{fspin} (a), we plot a free energy of mixing diagram $\Delta g_b^{mix}$ 
as a function of drug molar fraction $X_{d}$. In this diagram, the points 
$X_{d}^{1s}$, $X_{d}^{2s}$ are the solutions to 
\[ \frac{d^{2}\left(\Delta g_b^{mix}\right)}{dX_d^2} = 0,  \] 
and are referred to as the spinodal points. The region $(X_{d}^{1s}, X_{d}^{2s})$ is 
referred as the spinodal region, and for points $X_d$ in this region, we have
\[ \frac{d^{2}\left(\Delta g_b^{mix}\right)}{dX_d^2} < 0. \] 
Compositions $X_d$ in the spinodal region are unstable, and will split into two phases
characterized by the compositions $X_{d}^{1u}$ and $X_{d}^{2u}$ as shown in
Figure \ref{fspin} (a); see \cite{Porter:2009} for more details. The points 
$X_{d}^{1u}$, $X_{d}^{2u}$ are referred to as the binodal points, and are defined 
by the common tangent construction shown in Figure \ref{fspin} (a). The binodal and
spinodal points define the coexistence and spinodal curves, respectively, and these 
are plotted in the phase diagram shown in Figure \ref{fspin} (b). 

Using equation (\ref{eq:Gb}), we obtain 
\begin{equation}
\frac{d^{2} (\Delta g_b^{mix})}{dX_d^2} = RT \frac{q(X_d)}{(1 - (1 - 1/m)X_d)^{3}X_d(1 - X_d)}  
\label{eq:curvature}
\end{equation} 
where 
\begin{equation}
 q(X_d) = A X_d^2 + B X_d +1   \label{eq:quadratic} 
\end{equation} 
and where
\begin{equation}
 A = \frac{1}{m^{3}} -\frac{1}{m^{2}} - (1-2\chi_{dp})\frac{1}{m}+1, 
\hspace{0.5cm}B=\frac{1}{m^2}+(1-2\chi_{dp})\frac{1}{m} - 2. 
 \label{eq:AB}  
\end{equation} 
Hence there is a spinodal region with $d^{2}(\Delta g_b^{mix})/dX_d^2<0$ if $q(X_d)<0$ in 
this region. Inspecting (\ref{eq:quadratic}), we see that $q(X_d)$ can be negative 
if $q(X_d)=0$ has real roots, that is, if 
\[ B^{2} - 4A > 0,   \] 
and using (\ref{eq:AB}), this leads to 
\[   (2\chi_{dp} - (1+1/m))^{2} - 4/m  > 0 \] 
which holds true if  
\[  \chi_{dp} > \frac{1}{2}\left(1+\frac{1}{\sqrt{m}}\right)^{2} = 
     \frac{1}{2}\left(1+\sqrt{V_d/V_p}\right)^{2}.  \]
Hence, we have a spinodal interval if   
\begin{equation}
\chi_{dp} > \chi_{dp}^{c}(m) \label{eq:ineqsp}
\end{equation}  
where 
\begin{equation}
 \chi_{dp}^{c}(m) \equiv \frac{1}{2}\left(1+\frac{1}{\sqrt{m}} \right)^{2}, \label{eq:critchi}
\end{equation}  
and where $\chi_{dp}^{c}(m)$ is a critical value for the Flory-Huggins parameter. 
If (\ref{eq:ineqsp}) holds true, then there is a spinodal interval 
$(X_{d}^{1s},X_{d}^{1s})\subset [0,1]$ where
\begin{equation} 
X_{d}^{1s} = \frac{-B-\sqrt{B^{2}-4A}}{2A}, \hspace{0.3cm} 
X_{d}^{2s} = \frac{-B+\sqrt{B^{2}-4A}}{2A},    \label{eq:spinodal}
\end{equation}
and where $A$, $B$ are given in (\ref{eq:AB}). 

\vspace{0.3cm}

\noindent {\em The diffusion coefficient and spinodal decomposition}

\vspace{0.2cm}

Using (\ref{eq:effdiffN}) and (\ref{eq:curvature}), elementary calculations show that 
\begin{equation} 
D_{\mbox{\scriptsize eff}}(X_d) = M^{\dagger}(X_d) \frac{d^{2}(\Delta g_b^{mix})}{dX_d^{2}}   
\label{eq:diffeqn}
\end{equation}
where
\[  M^{\dagger}(X_d) = M_d X_d (1-X_d) \]
with $M_d=D_d/RT$, and where $M^{\dagger}(X_d)$ is a concentration-dependent drug mobility; 
see, for example, equation (3.6) of the paper \cite{Naumana:2001}. Hence, for $0<X_d<1$, it 
is clear from (\ref{eq:diffeqn}) that $d^{2}(\Delta g_b^{mix})/d X_d^2<0$ implies that    
\begin{equation}
D_{\mbox{\scriptsize eff}}(X_d) < 0.  \label{eq:uphilldiff} 
\end{equation}
Hence, an equivalent criterion for spinodal decomposition to occur is that there exist a 
region in $0\leq X_d\leq 1$ where $D_{\mbox{\scriptsize eff}}(X_d)<0$, that is, that 
there exist a region where drug diffusion is against the concentration gradient ({\em uphill 
diffusion}). 

\begin{figure}[ht] 
\begin{center}
\includegraphics[width=15.4cm,height=7.5cm]{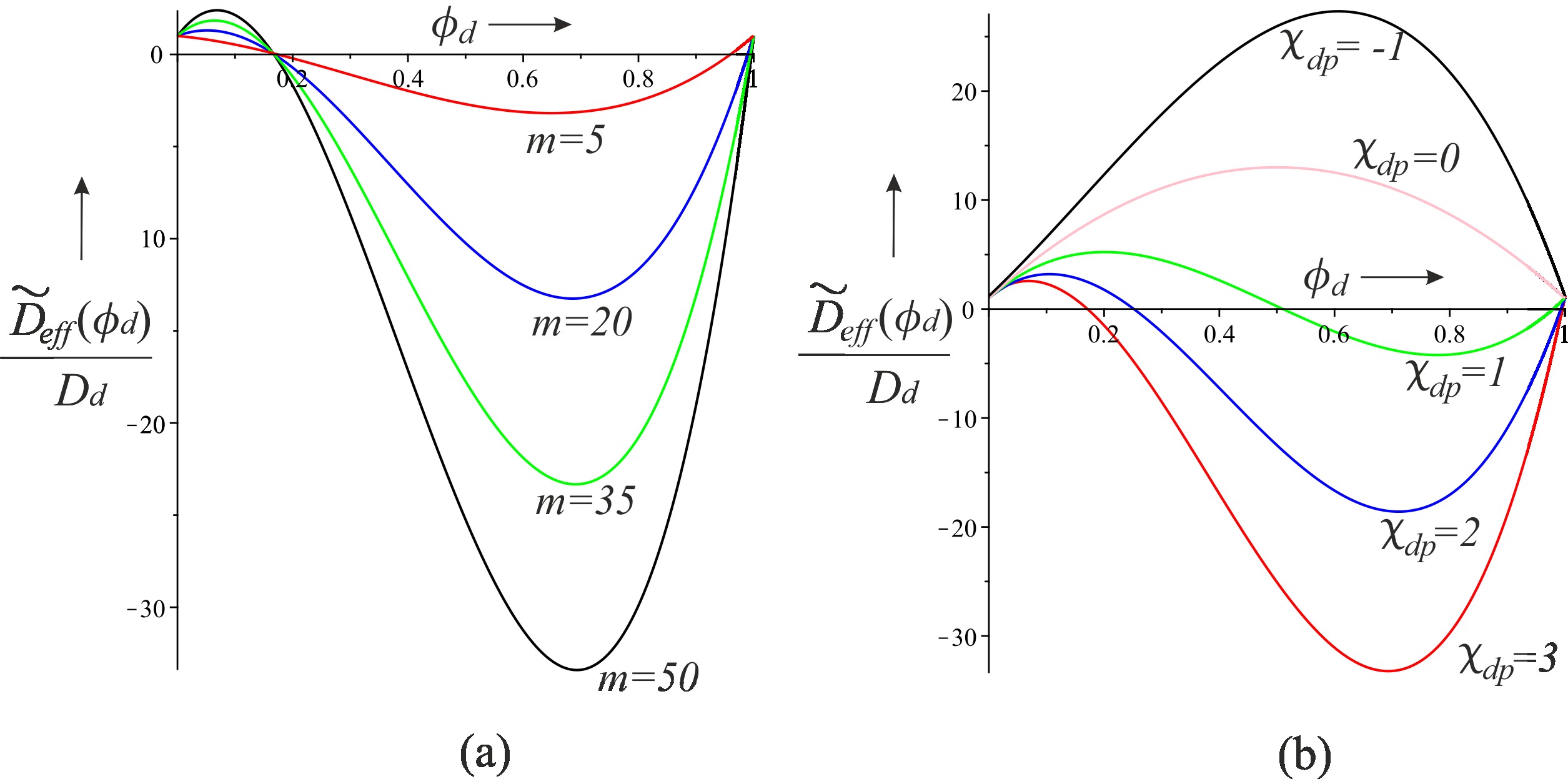}
\end{center}
\caption{Plots of the scaled effective diffusion coefficient for the drug in the polymer 
dispersion as a function of the drug volume fraction. Here positive values of the diffusion
coefficient correspond to standard drug diffusion down the concentration gradient, while 
negative values correspond to phase separation of the drug and the polymer, with  
larger negative values (in absolute terms) corresponding to more rapid phase separation. 
We have plotted the scaled drug diffusion coefficient for (a) the Flory-Huggins interaction 
parameter $\chi_{dp}=3$ and various values of the polymer chain length $m$, and, 
(b) polymer chain length $m=50$ and various values of the Flory-Huggins interaction 
parameter $\chi_{dp}$. See the main body of the text for further discussion.} 
\label{fig:Mobility} 
\end{figure}  

\section{Qualitative results and discussion} 

Although the model we have derived in the current study is quite general, and is not 
tied to any specific statistical model for a solid dispersion, the detailed results we shall
present in this section are for the Flory-Huggins case. 

\subsection{The effective diffusion coefficient for the drug in the dispersion}

From (\ref{eq:effdiffNphi}), the scaled effective diffusion coefficient for the drug in 
the dispersion is given by 
\begin{equation}
 \frac{\tilde{D}_{\mbox{\scriptsize eff}}(\phi_d)}{D_{d}}  
  =  (1 + (m-1)\phi_{d})\left( 1 + \left(\frac{1}{m} - 1\right)\phi_d  - 
       2\chi_{dp}\phi_{d}(1 - \phi_{d}) \right)   \label{eq:scaledeffdiffN}   
\end{equation}  
where we recall that $D_{d}$ is the temperature-dependent self-diffusion coefficient 
for the drug. Equation (\ref{eq:scaledeffdiffN}) is of particular value since it yields 
information on how the mobility of the drug in the dispersion depends on the polymer 
chain length, the dispersion composition, and the character of the drug-polymer 
interaction. 

In Figure \ref{fig:Mobility} (a), we have plotted (\ref{eq:scaledeffdiffN}) for the 
Flory-Huggins interaction parameter $\chi_{dp}=3$ (which is in the unstable regime)
and various values of the polymer chain length $m$. It should be emphasized that 
positive values for $\tilde{D}_{\mbox{\scriptsize eff}}$ correspond to standard drug 
diffusion down the concentration gradient, while negative values correspond to unstable 
regimes where phase separation of the drug and polymer can occur. In Figure 3 (a), it is 
clear that if the drug loading $\phi_{d}$ is sufficiently low, then 
$\tilde{D}_{\mbox{\scriptsize eff}}>0$ and the solid dispersion is stable. However,
for larger (and more realistic) drug loadings, $\tilde{D}_{\mbox{\scriptsize eff}}<0$, 
and the system is unstable. It is interesting to note that the system becomes more
unstable as the length of the polymer chains increase.  

It is also clear from the curves in Figure \ref{fig:Mobility} (a) that the relationship 
between the initial drug loading in the dispersion and the initial rate of phase separation 
is not altogether obvious. It is {\em not} necessarily the case that increasing drug 
loading corresponds to increasing initial dispersion instability. Rather, there is in fact a 
well defined {\em  worst choice} for the initial drug loading from the point of view 
of stability in the initial stages. This worst choice corresponds to the minima of the 
curves displayed in Figure \ref{fig:Mobility} (a), since these minima correspond to the
fastest rates of phase separation.  For $m\gg 1$, the minimum of 
$\tilde{D}_{\mbox{\scriptsize eff}}(\phi_{d})$ occurs at 
\begin{equation} 
\phi_{d}^{min} \approx 
\frac{1+2\chi_{dp}+\sqrt{(1+2\chi_{dp})^{2}-6\chi_{dp}}}{6\chi_{dp}}.    
\label{eq:worstcase} 
\end{equation} 
These theoretical results predict that choosing initial drug loadings $\phi_{d}$ 
above or below $\phi_{d}^{min}$ should lead to improved dispersion stability 
in the initial stages. For $m,\chi_{dp}\gg 1$, we have $\phi_{d}^{min}\approx 0.67$.

In Figure \ref{fig:Mobility} (b), we plot (\ref{eq:scaledeffdiffN}) for the fixed 
polymer length $m=50$, and various values of the Flory-Huggins interaction 
parameter $\chi_{dp}$. For $m=50$, the critical value for $\chi_{dp}$ is given 
by $\chi_{dp}^{c}\approx 0.5707$ (see equation (\ref{eq:critchi})). Recall that 
for $\chi_{dp}<\chi_{dp}^{c}$, the system is stable for all drug loadings 
$\phi_{d}$, and that for $\chi_{dp}>\chi_{dp}^{c}$, there is a regime of 
unstable drug loadings. This is borne out by the curves displayed in 
Figure \ref{fig:Mobility} (b).  These curves predict that the system becomes 
more unstable with increasing values of $\chi_{dp}$, and this is as expected 
given the dependence of $\chi_{dp}$ on the interaction energies - see equation 
(\ref{eq:chienergies}).

\begin{figure}[ht] 
\begin{center}
\includegraphics[width=14.0cm,height=8.0cm]{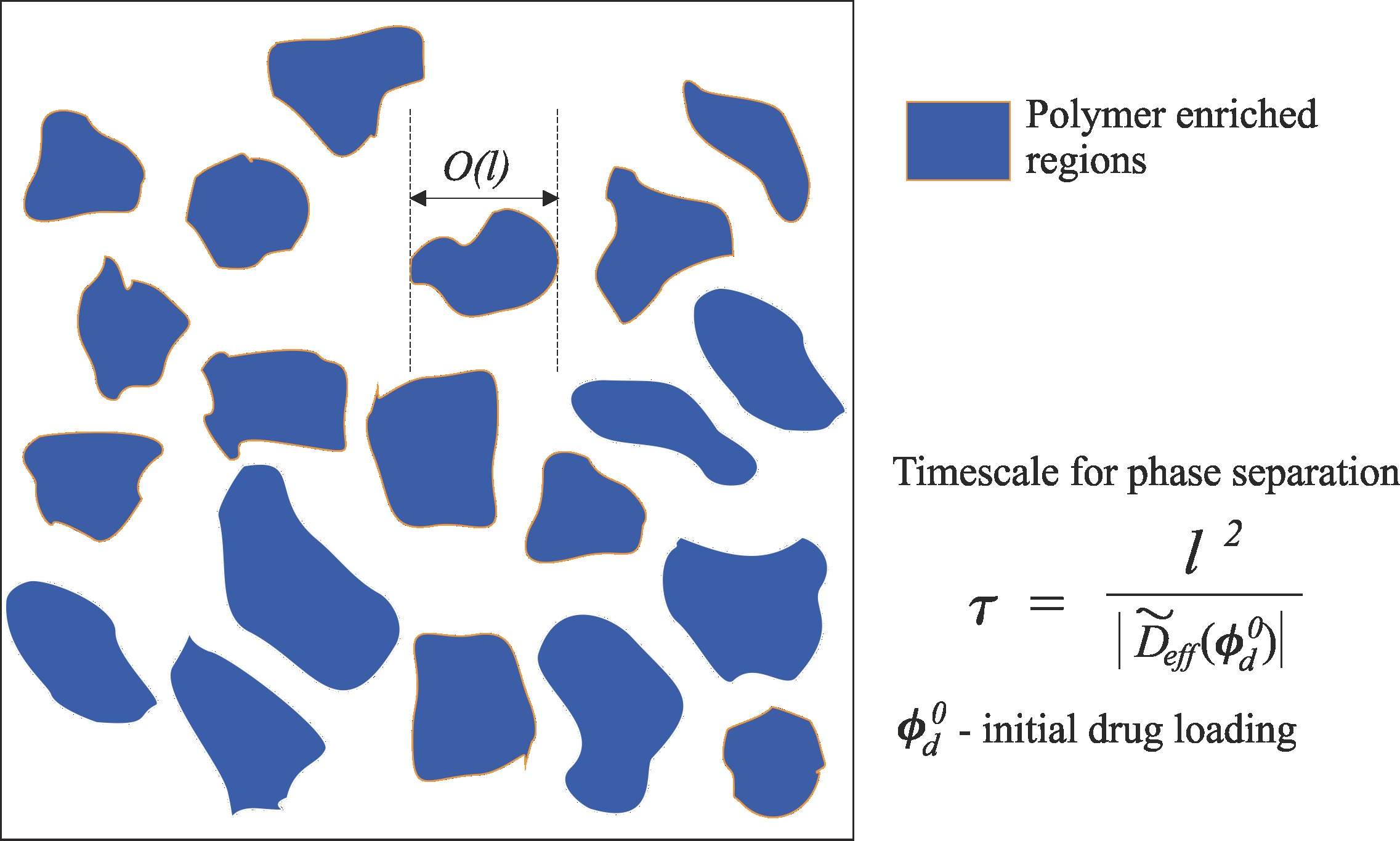}
\end{center}
\caption{Schematic of a phase separating solid dispersion where polymer-rich 
regions with characteristic lengthscale $l$ have formed. A formula for the 
timescale of evolution of such a dispersion is given in the main body of the 
text; see equation (\ref{eq:diffusiontimescale}).} 
\label{fig:tps} 
\end{figure} 

\subsection{Timescale for phase separation in a solid dispersion}

In Figure \ref{fig:tps}, we give a schematic of a phase separating solid dispersion 
where polymer-rich regions have formed. The characteristic lengthscale of these 
regions is denoted by $l$. In order for such regions to form, the drug must have 
diffused away over a lengthscale of order $l$, and the timescale over which this 
diffusion occurs is estimated by (see (\ref{eq:effdiffNphi})) 
\begin{equation}
\tau = \frac{l^{2}} {\vert \tilde{D}_{\mbox{\scriptsize eff}} (\phi_d^0) \vert } =
\frac{l^{2}}{D_{d}(T)}\frac{1}
{\vert (1 + (m - 1)\phi_d^0)[ 1 + (1/m - 1)\phi_d^0 - 
2\chi_{dp}(T)\phi_{d}^{0}(1-\phi_{d}^{0})] \vert }   \label{eq:diffusiontimescale}
\end{equation} 
where $\phi_d^0$ is the initial uniform volume fraction of the drug in the dispersion,
and $T$ is a representative storage temperature. It should be emphasized that this 
formula is just an estimate since, in reality, the drug volume fraction evolves in space 
and time. Hence, (\ref{eq:diffusiontimescale}) should only be used as a rough rule of 
thumb. In Section \ref{numericss}, we evaluate this formula by comparing it with 
detailed numerical results, and satisfactory agreement is generally found.  

Equation (\ref{eq:diffusiontimescale}) may, in appropriate circumstances, be used to 
estimate the shelf life of a solid dispersion product. To see this, suppose that $l$ 
denotes the largest acceptable size for polymer-rich domains (or drug-rich domains) 
in the product.  Then, since $\tau$ estimates the timescale for these regions to 
form, it also estimates the timescale for the shelf life of the product. However, care 
should be taken when using (\ref{eq:diffusiontimescale}) since, apart from the 
fact that is based on a fixed value of $\phi_{d}$, it also incorporates a number of 
significant assumptions - for example, it assumes that the dispersion is perfectly dry, 
and that Flory-Huggins theory is an appropriate statistical model for the system. 
  
\subsection{Criteria for a stable solid dispersion}

Although the drug loading in real solid dispersions is typically high and in the unstable 
regime, it is nevertheless worthwhile specifying conditions under which the stability of 
the dispersion is guaranteed. The results we display here are based on the discussion 
given in Section \ref{sec:phaseseparation}. For $\chi_{dp}<\chi_{dp}^{c}$ where
$\chi_{dp}^{c}=\frac{1}{2}\left(1 + 1/\sqrt{m}\right)^{2}$, the system is stable 
irrespective of the choice of the uniform initial drug load $\phi_{d}^{0}$. For 
$\chi_{dp}>\chi_{dp}^{c}$, the dispersion is unstable if the initial drug loading 
$\phi_{d}^{0}$ is chosen  in the interval $(\phi_{d}^{-},\phi_{d}^{+})$, but stable
if chosen in either of the intervals $(0,\phi_{d}^{-})$ or $(\phi_{d}^{+},1)$, where
\[ \phi_{d}^{\pm} = \frac{1}{2}\left\{ 1 + \frac{1}{2\chi_{dp}}
       \left(1-\frac{1}{m}\right)\pm 
       \sqrt{ \left[ 1 + \frac{1}{2\chi_{dp}}\left(1-\frac{1}{m}\right)\right]^{2} 
        - \frac{2}{\chi_{dp}} } \right\}.      \] 
These results are based on the bulk free energy only, and do not take account of 
interfacial energy. However, the interfacial energy can be readily incorporated into the 
analysis, and this is discussed in Appendix A. 
 
\section{Numerical results and discussion \label{numericss}}

\subsection{The numerical method  \label{sec:method} }

For the purposes of numerical calculations, we take the integration domain to 
be the square region $\Omega=\{(x,y) \vert \hspace{0.1cm} 0<x<L,0<y<L\}$ 
with boundary $\partial\Omega$.  The governing equation to be solved is defined
by the equations  (\ref{eq:eqNd}), (\ref{eq:defpsi}) and (\ref{eq:eqchi}). The 
boundary conditions are given by
\begin{equation}
  \nabla\psi\cdot {\bf n} = 0 \hspace{0.3cm}\mbox{ and }\hspace{0.3cm}
 \nabla X_{d}\cdot {\bf n} = 0 \hspace{0.3cm}\mbox{ on }\hspace{0.1cm}
 \partial\Omega,   \label{eq:bcs2d2} 
\end{equation}
and the initial condition takes the form (\ref{eq:icform}).  The boundary 
conditions (\ref{eq:bcs2d2}) are equivalent to those given in (\ref{eq:bcs2d}). 
The governing equation was numerically integrated using the finite 
element package COMSOL Multiphysics.  A mesh sensitivity analysis was performed 
to investigate the influence of the size of the mesh on the results. The solution was 
assumed to be mesh independent when there was less than 1\% difference in the 
mole fraction of drug between successive refinements. The final mesh used in the 
simulations was triangular and consisted of 7553 vertices and 14796 triangles. The 
numerical solutions all conserved the total mass of drug in the system to within 1\%. 

\subsection{Parameter values \label{sec:parvalues} }

We consider parameter values that are appropriate for a solid dispersion consisting 
of the drug Felodipine (FD) and the polymeric excipient HPMCAS. Felodipine is a 
calcium channel blocker that is commonly used to treat blood pressure. For this 
system, the Flory-Huggins interaction parameter is given as a function of temperature 
by (see \cite{Tian2:2013})
\begin{equation} 
   \chi_{dp}(T) = - 18.767 + \frac{7830.4}{T}.  \label{eq:chidpT}
\end{equation}    
Using data taken from \cite{Tian2:2013}, the molar volume for FD is 
$V_d = 300.19$ cm$^{3}$/mol and the molar volume of HPMCAS is 
$V_{p}=14007.78$ cm$^{3}$/mol, so that
\[  m =  \frac{V_{p}}{V_{d}} = \frac{14007.78}{300.19} \approx 46.6630.  \] 
From (\ref{eq:critchi}), the critical value for the interaction parameter below which phase
separation cannot occur is given by
\[ \chi_{dp}^{c}(m) = \frac{1}{2}\left(1+\frac{1}{\sqrt{m}}\right)^{2} = 0.6571.  \] 
The self-diffusion coefficient for Felodipine was estimated in \cite{Gerges:2015} 
(Chapter 4, page 133) to be 
\begin{equation}
 D_{d}(T) = \exp(-A_{1}) \exp\left(-\frac{A_{2}}{T}
  \exp\left(\frac{A_{3}}{T} \right)  \right)  
  \hspace{0.2cm} \mbox{m}^{2}\mbox{s}^{-1} \label{eq:DpT}
\end{equation}
where $A_{1}=18.03$, $A_{2}=445.84$ K, $A_{3}=874.81$ K. Some illustrative 
values for the diffusion coefficients  and the Flory-Huggins interaction parameter are 
displayed in Table 1. 

For the numerical simulations displayed in the current study, we take the size 
of the square domain to be given by $L=2$ mm. The thickness of the interfacial
regions is dictated by the parameter $\delta_d$, and here we chose the value
 $\delta_d = L/50 = 4 \times 10^{-5}$ m. We illustrate how the initial conditions 
were specified by considering a particular case. We consider the case where the 
initial weight fraction of drug is 80\%. This means that the initial weight of FD 
divided by the weight of FD plus the weight of HPMCAS is 0.8. This corresponds 
to an initial molar drug fraction of $X_{d} = 0.9947$. More 
precisely, we choose the initial molar fraction of drug to be a small random 
perturbation about this level given by 
\[ X_{d}(x,y,t=0) = 0.9947 (1 + rnd(x,y)) \]
where $rnd(x,y)$  is a normally distributed random function with a mean value 
of zero and a standard deviation of $10^{-5}$. The standard deviation 
for all of the initial conditions was taken to be $10^{-5}$, with one exception - the 
numerical results displayed in Figure \ref{fig:hme} took the larger value $0.007$ to 
simulate a coarse initial mixture in a hot melt extruder.

\begin{figure}[ht] 
\begin{center}
\includegraphics[width=11cm,height=7cm]{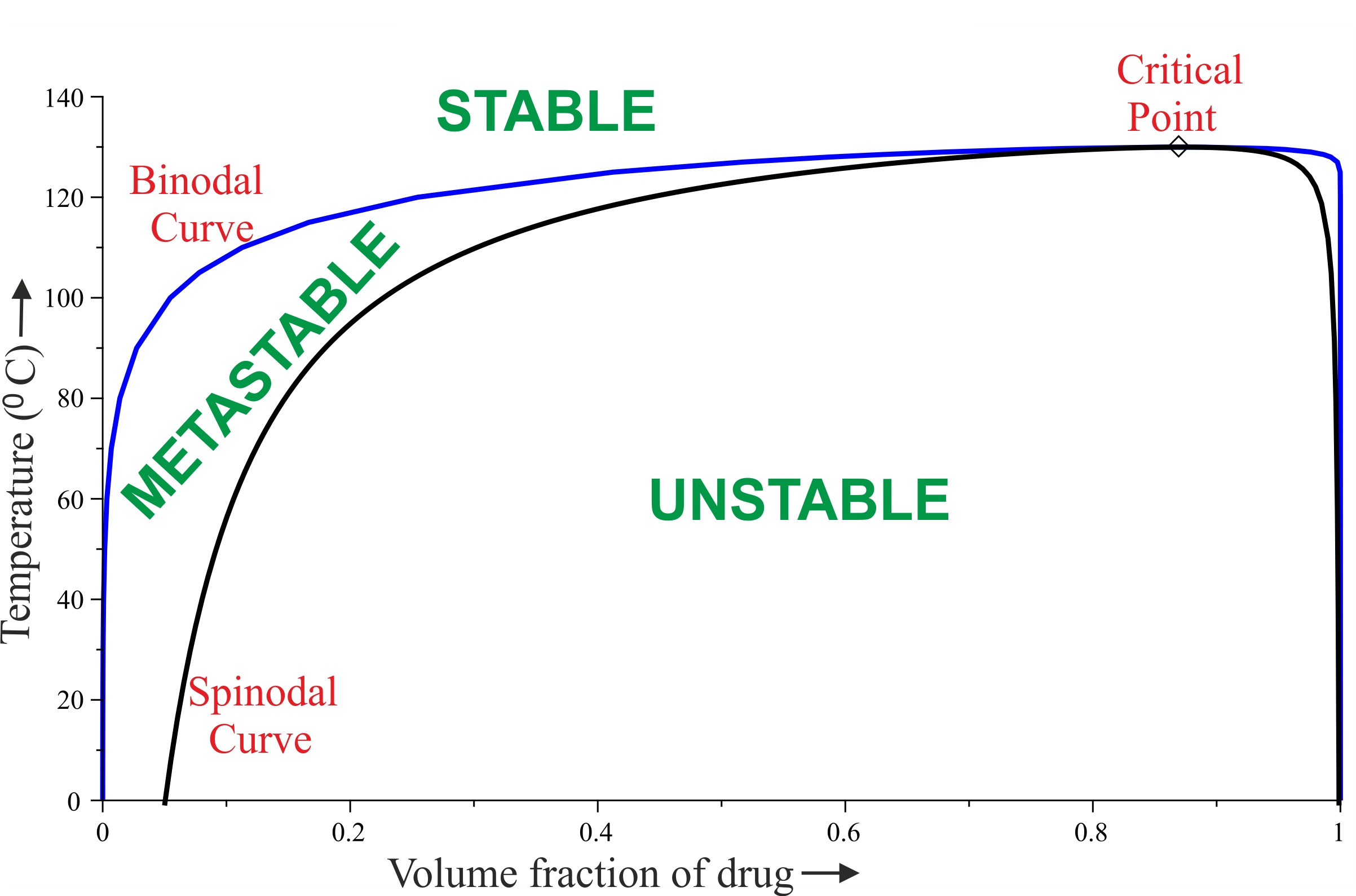}
\end{center}  
\caption{The phase diagram for the Felodipine/HPMCAS system.} 
\label{fig:PhaseDiagram} 
\end{figure}

\begin{figure}[ht] 
\begin{center}
\includegraphics[width=11cm,height=9.5cm]{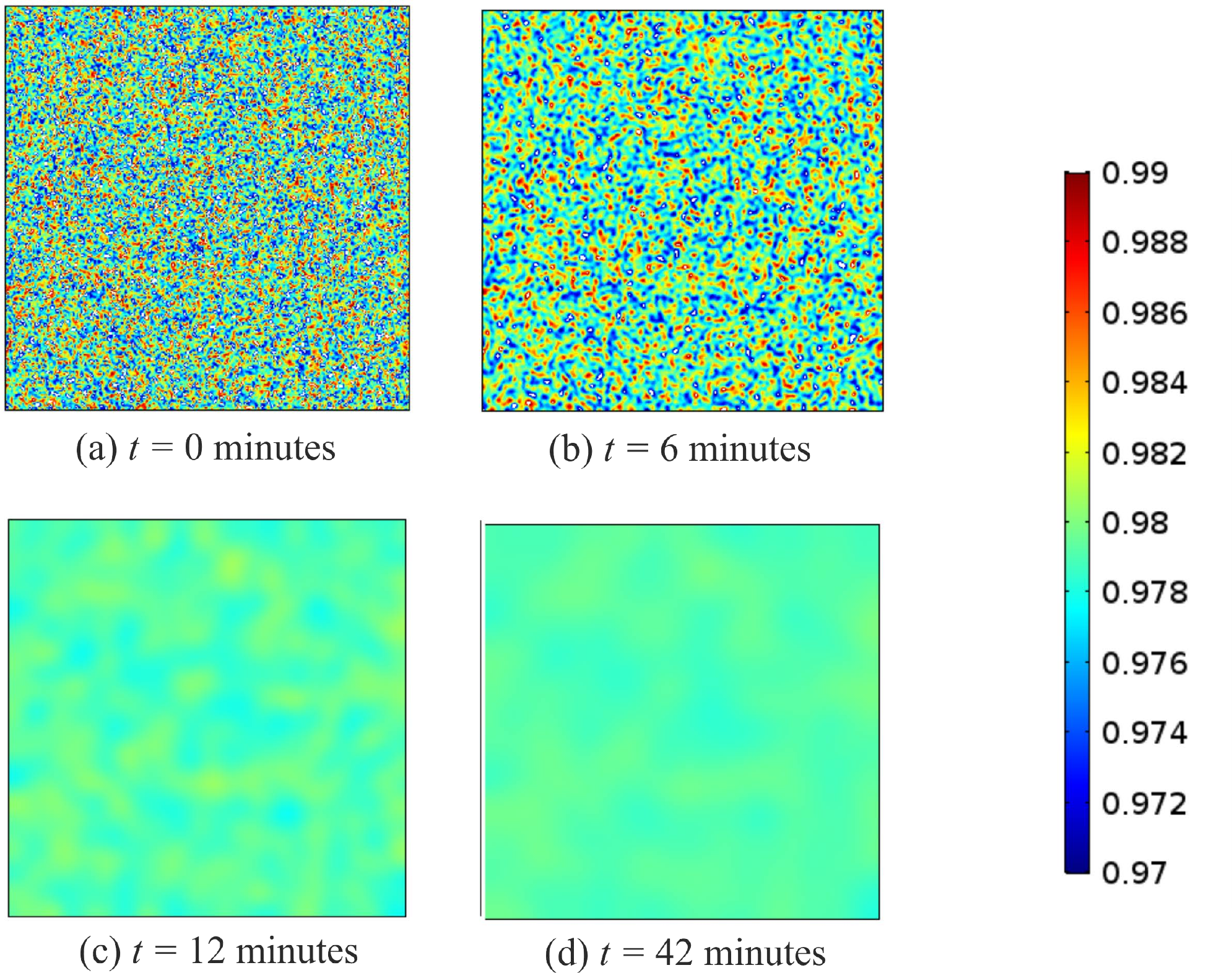}
\end{center}  
\caption{Numerical simulation of the behaviour of a solid dispersion
             in a hot melt extruder. The simulations here are for a FD/HPMCAS solid dispersion 
             and were obtained by numerically integrating the initial boudary value problem 
             defined in  Section \ref{sec:method}. The colours correspond to different mole 
             fractions of the drug as defined by the colour bar. The weight fraction 
             of drug here is 50\%, and the other parameter values can be found in 
             Section \ref{sec:parvalues}  and Section \ref{sec:numress}. These simulations 
             represent a succesful extrusion where an initially coarse mixture is heated and 
             then cooled to form a well-mixed dispersion.} 
\label{fig:hme} 
\end{figure}

In Figure \ref{fig:PhaseDiagram}, we plot the phase diagram for the 
Felodipine/HPMCAS system. All that is required to calculate the phase diagram
here is a knowledge of the Flory-Huggins interaction parameter, and this has been
given in  (\ref{eq:chidpT}). The spinodal curve $T_{s}(\phi_d)$ is obtained by setting
$\tilde{D}_{\mbox{\scriptsize eff}}(\phi_d)=0$ in (\ref{eq:scaledeffdiffN}) to obtain 
\begin{equation}
1+\left(\frac{1}{m}-1\right)\phi_{d} - 2\chi_{dp}\phi_{d}(1 - \phi_{d}) = 0, 
\label{eq:eqfspin}
\end{equation}  
where $\chi_{dp}$ is given by (\ref{eq:xhidpT}). Solving (\ref{eq:eqfspin}) for 
$T$ gives the spinodal curve 
\begin{equation*}
T_{s}(\phi_{d}) = \frac{2\beta\phi_{d} (1 - \phi_{d})}{1+\left(\frac{1}{m}-1\right)\phi_{d}
 - 2\alpha\phi_{d}(1 - \phi_{d})}
\end{equation*}
where $\alpha = -18.767$, $\beta = 7830.4$, $m=46.6630$ for the Felodipine/HPMCAS 
system. The binodal curve was estimated numerically. The calculation involved 
simultaneously solving the pair of equations $\mu_d^b (X_d^{1u} )=\mu_d^b (X_d^{2u} )$ 
and $\mu_p^b (X_d^{1u} )=\mu_p^b (X_d^{2u})$ for the binodal points $X_d^{1u}$, 
$X_d^{2u}$ with $X_d^{1u}<X_d^{2u}$. This was implemented using the the {\tt fsolve} 
command in MAPLE. 
\begin{table}
\begin{center}
\begin{tabular}{ |c|c|c|c|}  
\hline
$T$ ($^{\mbox{o}}$C) & $\chi_{dp}(T)$ & $D_{d}(T)$ (m$^{2}$s$^{-1}$)  
& $D_{\mbox{\scriptsize eff}}(X_{a})$ (m$^{2}$s$^{-1}$) \\ 
\hline
$ 40 $ & 6.2383 & $1.1661\times 10^{-18}$  & $8.8605\times 10^{-17}$  \\  
$ 50 $ & 5.4645 & $1.5494\times 10^{-17}$ & $1.0113\times 10^{-15}$  \\  
$ 60 $ & 4.7371 & $1.3787\times 10^{-16}$ & $7.6107\times 10^{-15}$  \\  
$ 75 $ & 3.7245 & $2.0297\times 10^{-15}$ & $8.3587\times 10^{-14}$  \\ 
$ 90 $ & 2.7954 & $1.7356\times 10^{-14}$ & $4.9151\times 10^{-13}$  \\ 
$ 100 $ & 2.2176 & $5.7436\times 10^{-14}$ & $1.1670\times 10^{-12}$ \\ 
$ 110 $ & 1.6699 & $1.6336\times 10^{-13}$ & $2.0806\times 10^{-12}$ \\ 
$ 120 $ & 1.1501 & $4.0902\times 10^{-13}$ & $2.2657\times 10^{-12}$ \\ 
\hline
\end{tabular}
\caption{Illustrative values for some of the parameters of the FD/HPMCAS 
system at various temperatures. Here the initial weight fraction of drug is 70\%,
which corresponds to an initial drug molar fraction $X_{d}\approx 0.9909$.}   
\end{center}
\label{ta:chi}
\end{table}

\subsection{Numerical results  \label{sec:numress}}

The first numerical calculations we display simulate the hot melt mixing process 
for a Felodipine/HPMCAS system. To implement this, we used a time dependent temperature 
profile that treats the case where the mixture begins at 25$^{\mbox{o}}$C and rises in temperature at 
a rate of 10$^{\mbox{o}}$C per minute until it reaches 145$^{\mbox{o}}$C. We then assumed for simplicity 
that the melt cooled linearly back down to 25$^{\mbox{o}}$C over a period of 30 minutes. The results 
of the calculation are shown in Figure \ref{fig:hme}.  We see in Figure \ref{fig:hme} (a) that the 
initial drug/polymer mixture is quite coarse (badly mixed).  As the temperature rises, we see from 
Figure \ref{fig:hme} (b) that the mixture becomes increasingly homogenous. By the time the mixture 
has achieved its maximum temperature, Figure \ref{fig:hme} (c), it is quite well-mixed. 
Figure \ref{fig:hme} (d) shows that the amount of phase separation that occurs during the cooling 
process is insignificant. Hence, the numerical results shown here predict a successful hot melt 
extrusion process for the manufacture of a solid dispersion. We note that different heating and 
cooling regimes are easily simulated using the model.

\begin{figure}[ht] 
\begin{center}
\includegraphics[width=12.5cm,height=9.5cm]{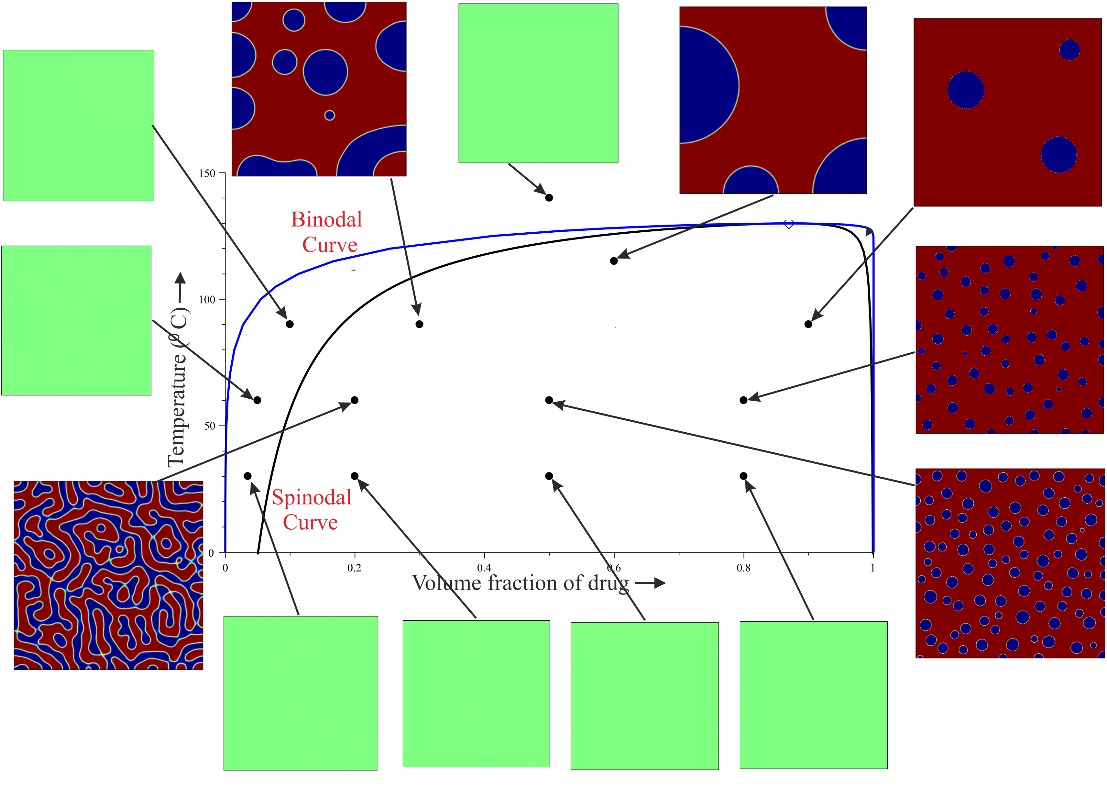}
\end{center}  
\caption{Numerical solutions superimposed on a phase diagram for the
                  Felodipine/HPMCAS system. The numerical solutions shown here are all 
                 for a time of six months. The parameter values used can be found in Section 
                 \ref{sec:parvalues}. The dark blue regions are polymer-enriched and the dark 
                 red regions are drug-enriched. The green panels correspond to cases where 
                 phase separation is not significant.} 
\label{fig:pdns} 
\end{figure}

In Figure \ref{fig:pdns}, we superimpose numerical solutions on a phase diagram 
for the Felodipine/HPMCAS system. Each of these solutions corresponds to an evolution time 
of 6 months, with the dispersion mixture beginning from an initially approximate uniform state. 
We see that there is no significant phase separation for the 30$^{\mbox{o}}$C degree cases, 
and for the cases in the metastable and stable regions. This predicts that Felodipine/HPMCAS
systems should not suffer considerable phase separation under normal storage temperatures.
However, we should caution that we are modelling the case of zero relative humidity here. 
For the cases that do exhibit significant phase separation, we note a coarser separation morphology 
for higher temperatures. In these figures, dark red corresponds to drug-enriched domains (relative 
to the initial concentration) while dark blue correspond to polymer-enriched domains. We also
note the occurrence of polymer droplets and strings - we return to this issue below. 

\begin{figure}[p]
\centering 
\vspace{-27mm}
\begin{minipage}[b]{0.90\linewidth}
\begin{tabular}{ccc}
$T=60^{\circ}C $& $T=90^{\circ} C$ & $T=120^{\circ} C$ \\
\includegraphics[width=35mm]{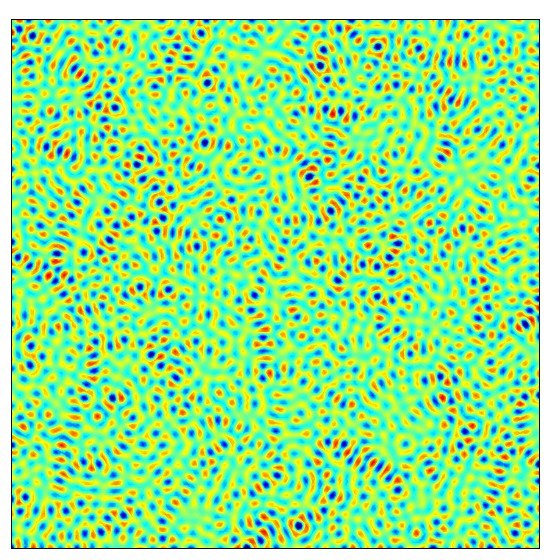} & 
\includegraphics[width=35mm]{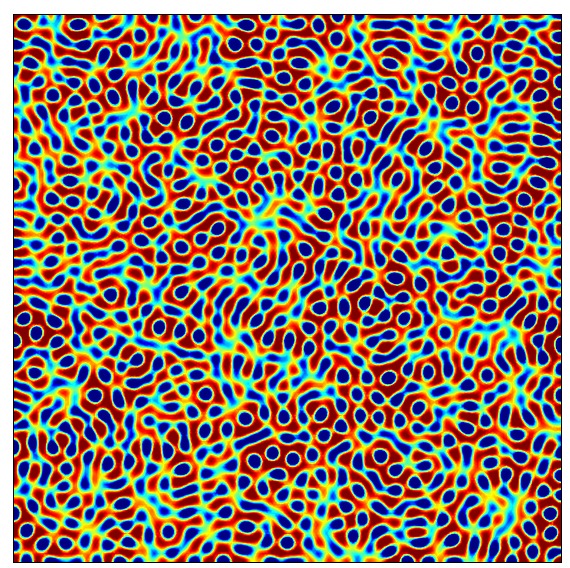} &
\includegraphics[width=35mm]{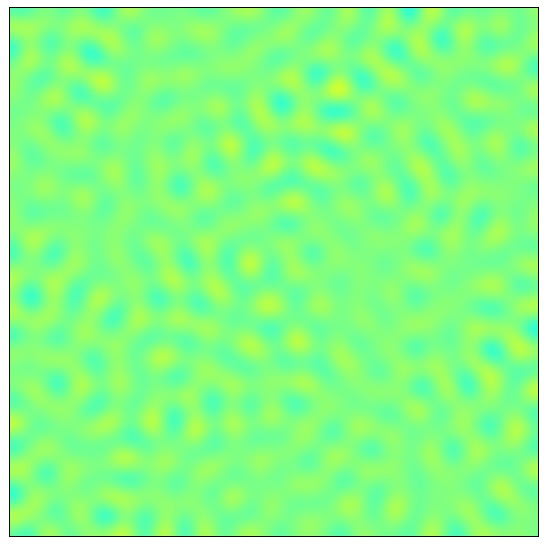} \\
\vspace{-7mm} \\
{\scriptsize 1 day} & {\scriptsize 1 hour} & {\scriptsize  0.5 hours}\\
\vspace{-10mm} \\
& & \\
\includegraphics[width=35mm]{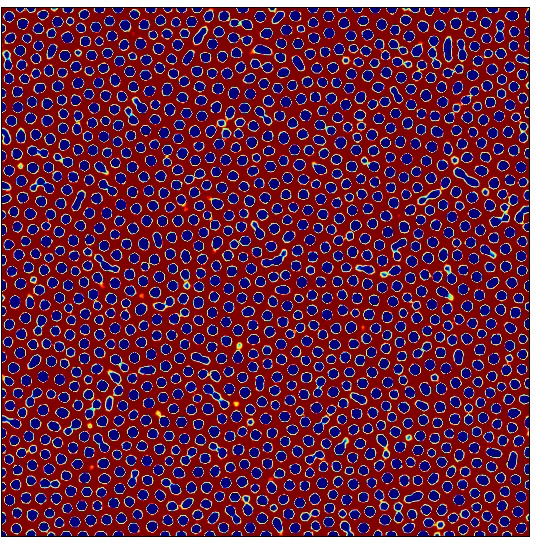}&
\includegraphics[width=35mm]{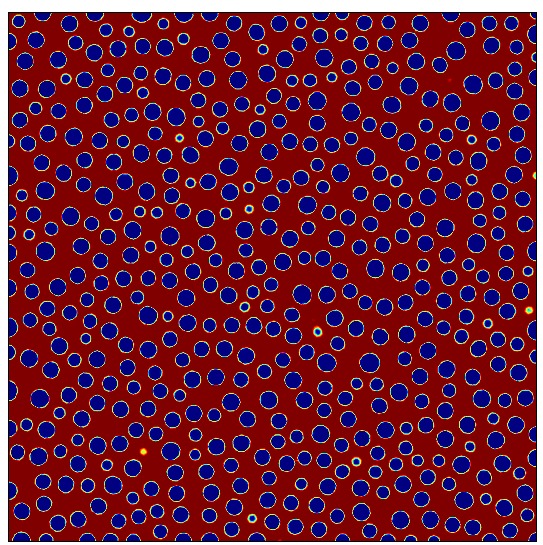}&
\includegraphics[width=35mm]{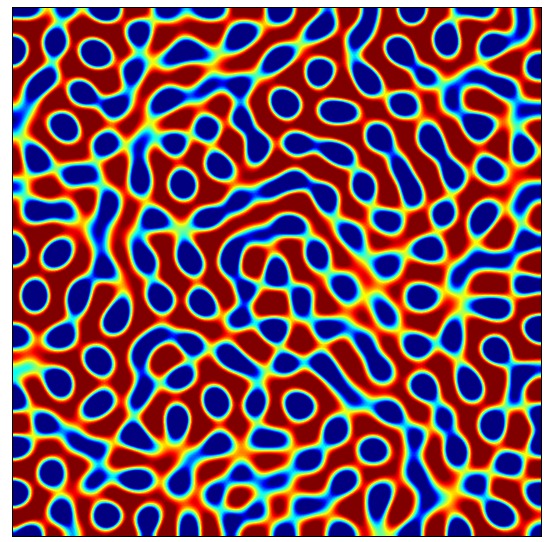}\\
\vspace{-7mm} \\
{\scriptsize 2 days} & {\scriptsize  3 hours} & {\scriptsize  1 hour}\\
\vspace{-10mm} \\
& & \\
\includegraphics[width=35mm]{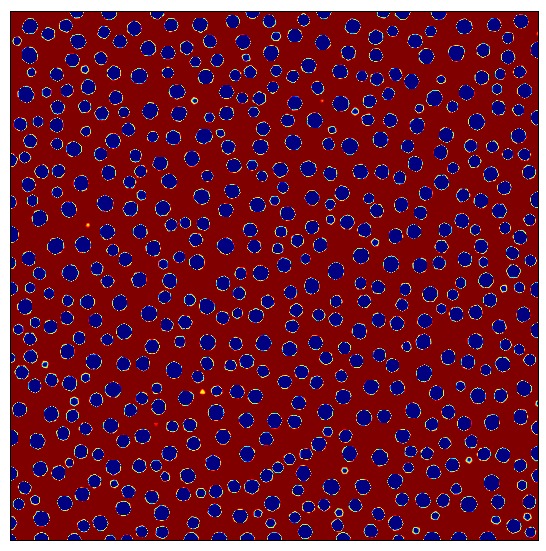}&
\includegraphics[width=35mm]{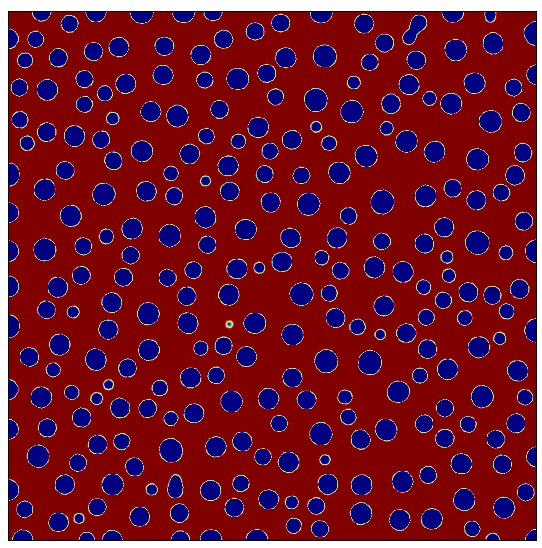}&
\includegraphics[width=35mm]{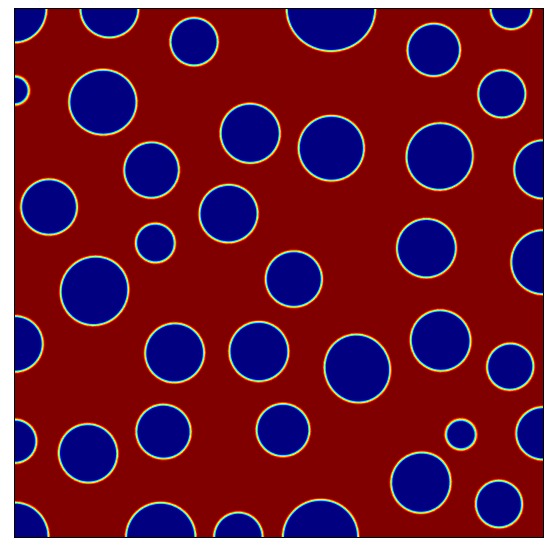}\\
\vspace{-7mm} \\
{\scriptsize 1 week} & {\scriptsize 4 hours} & {\scriptsize  1 day} \\
\vspace{-10mm} \\
& & \\
\includegraphics[width=35mm]{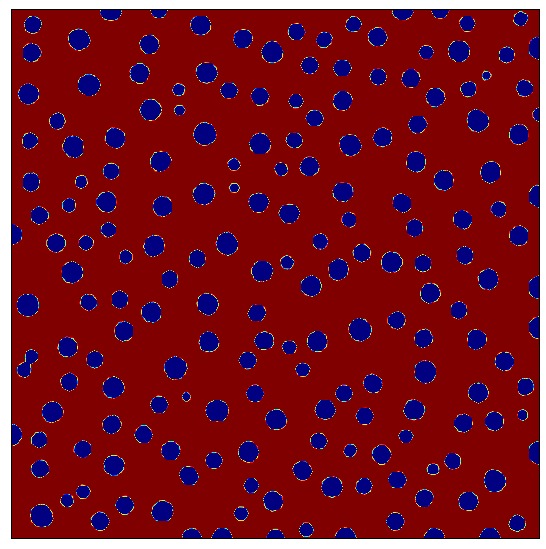}&
\includegraphics[width=35mm]{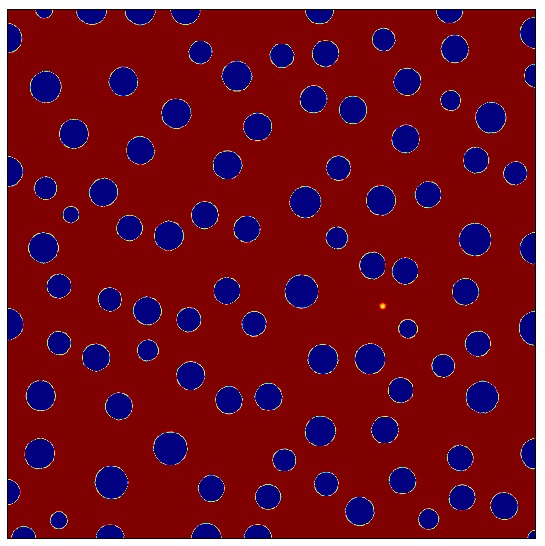}&
\includegraphics[width=35mm]{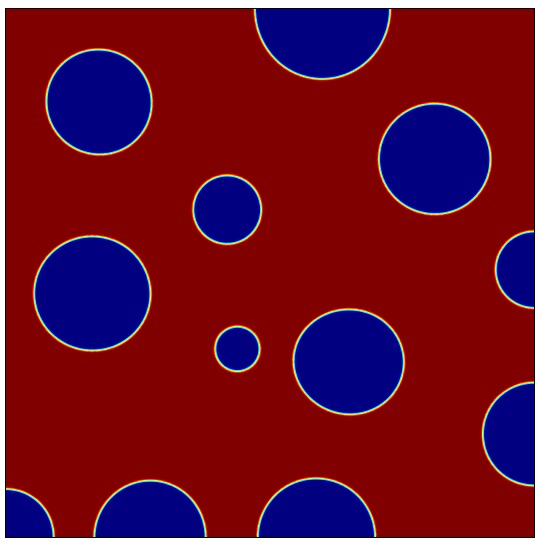}\\
\vspace{-7mm} \\
{\scriptsize 1 month} & {\scriptsize 1 day} & {\scriptsize 1 week}\\
\vspace{-10mm} \\
& & \\
\includegraphics[width=35mm]{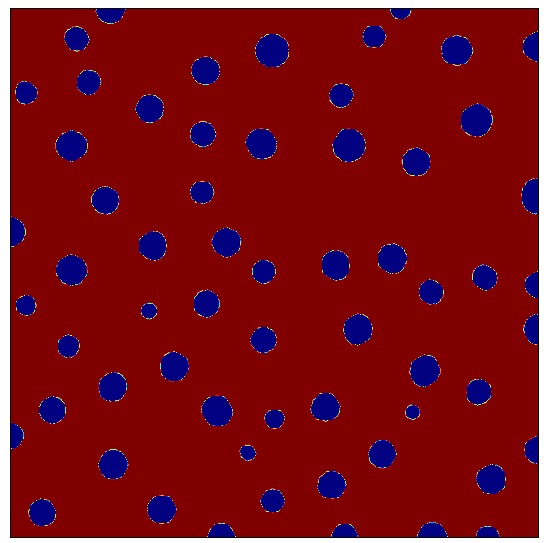}&
\includegraphics[width=35mm]{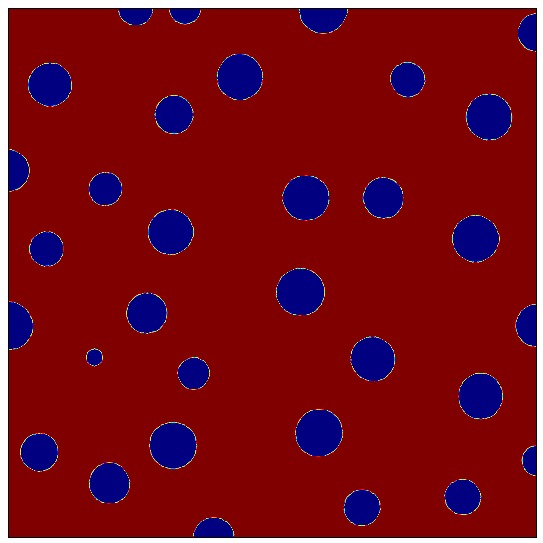}&
\includegraphics[width=35mm]{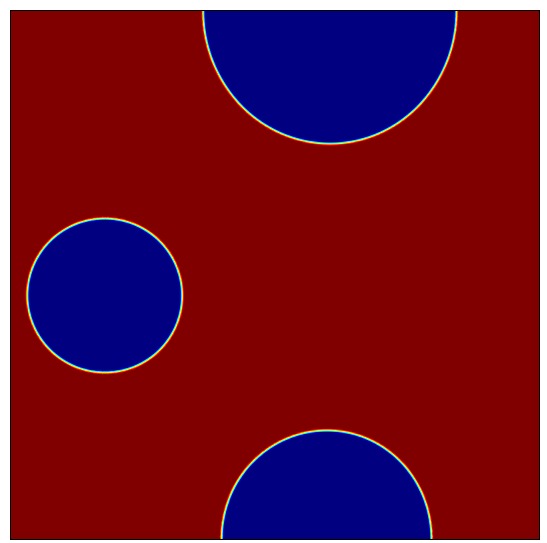}\\
\vspace{-7mm} \\
{\scriptsize 6 months} & {\scriptsize  1 week} & {\scriptsize 2 months}\\
\end{tabular}
\end{minipage}
\hspace{-20mm}
\begin{minipage}[b]{0.1\linewidth}
\begin{tabular}{c}
\\ \\ \\
\includegraphics[width=25mm]{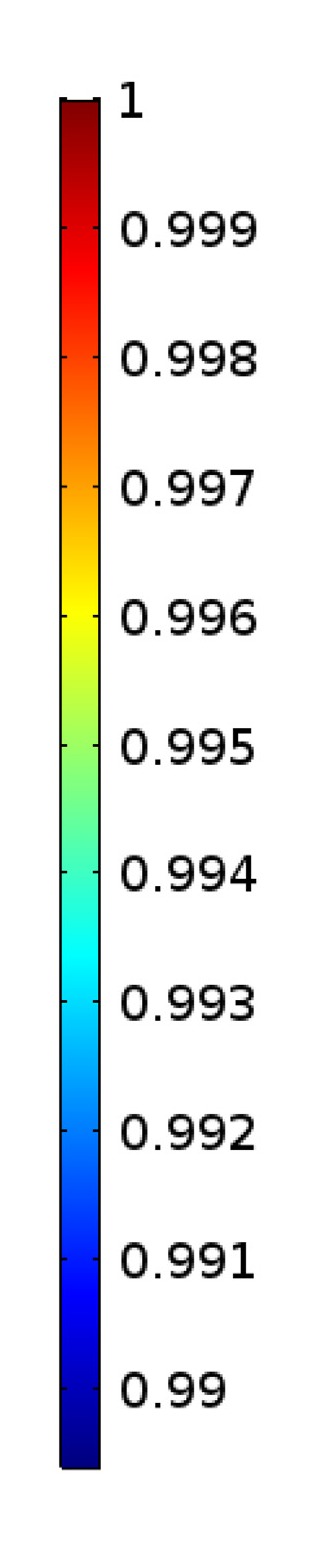} \\
\end{tabular}
\end{minipage}
\caption{Simulations of a FD/HPMCAS solid dispersion obtained by
             numerically integrating the initial boudary value problem defined in
             Section \ref{sec:method}. The colours correspond to different mole 
             fractions of the drug as defined by the colour bar. The weight fraction 
             of drug here is 80\%, and the other parameter values can be found in 
             Section \ref{sec:parvalues}. In the above frame of figures, each column
             corresponds to a different temperature, and reading a column from top 
             to bottom corresponds to increasing time for the dispersion for the 
             given temperature.}
\vspace{-10mm}
\label{fig:results80}
\end{figure}

\begin{figure}[p]
\centering 
\vspace{-27mm}
\begin{minipage}[b]{0.9\linewidth}
\begin{tabular}{ccc}
$T=60^{\circ}C $& $T=90^{\circ} C $& $T=120^{\circ} C $\\
\includegraphics[width=35mm]{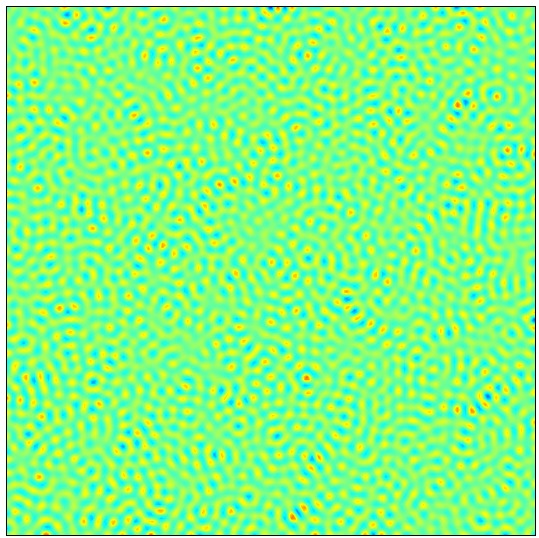} & 
\includegraphics[width=35mm]{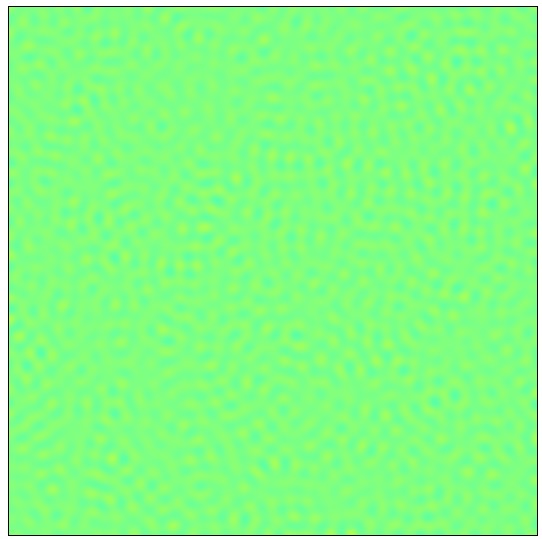} &
\includegraphics[width=35mm]{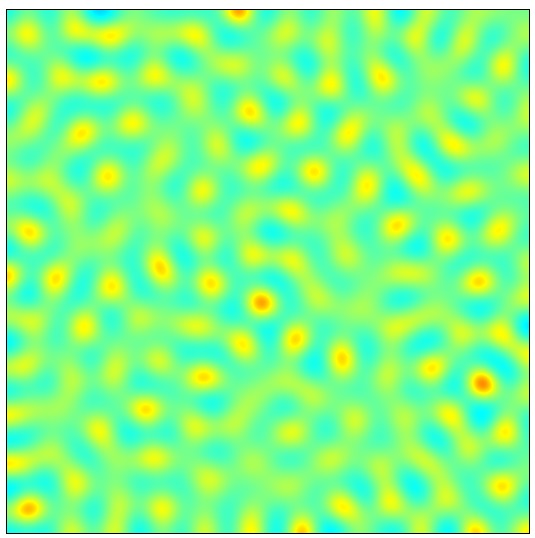} \\
\vspace{-7mm} \\
{\scriptsize 1 day} & {\scriptsize 0.5 hours} & {\scriptsize 1.5 hours}\\
\vspace{-10mm} \\
& & \\
\includegraphics[width=35mm]{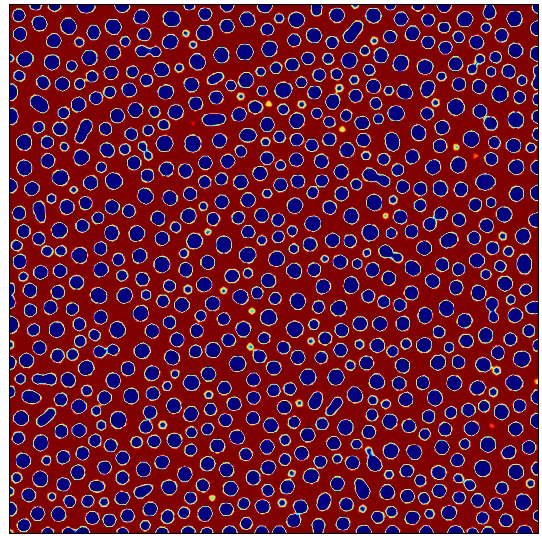}&
\includegraphics[width=35mm]{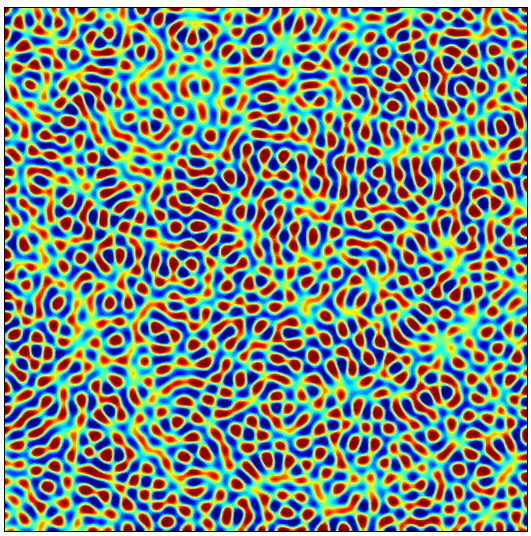}&
\includegraphics[width=35mm]{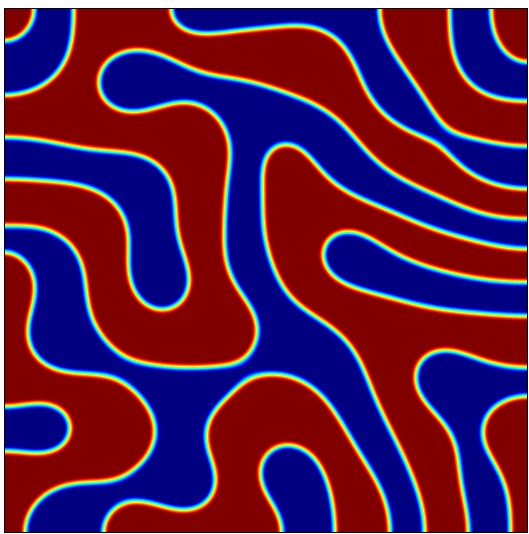}\\
\vspace{-7mm} \\
{\scriptsize 1 week} & {\scriptsize 1 hour} & {\scriptsize 1 day}\\
\vspace{-10mm} \\
& & \\
\includegraphics[width=35mm]{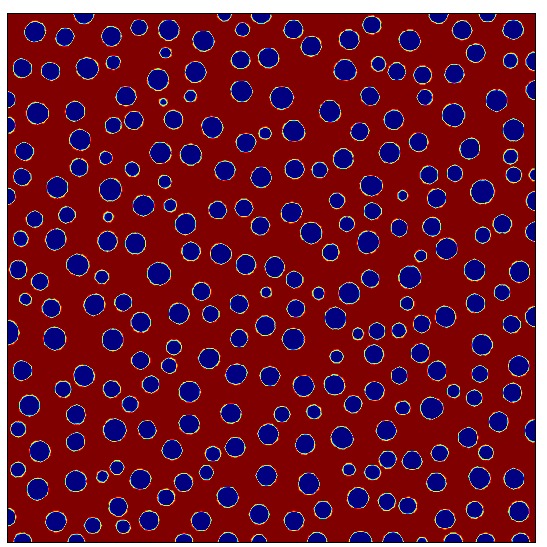}&
\includegraphics[width=35mm]{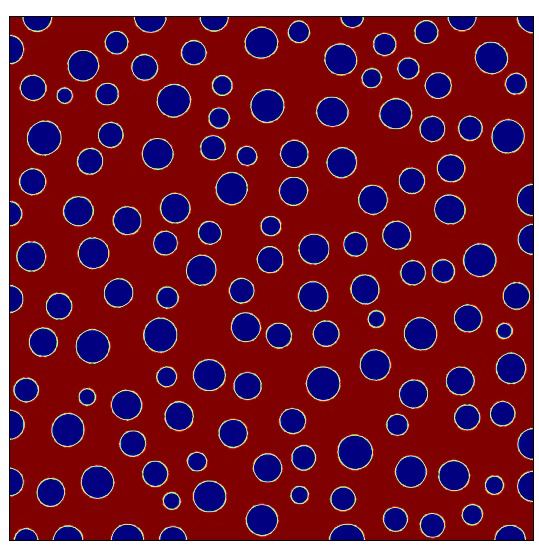}&
\includegraphics[width=35mm]{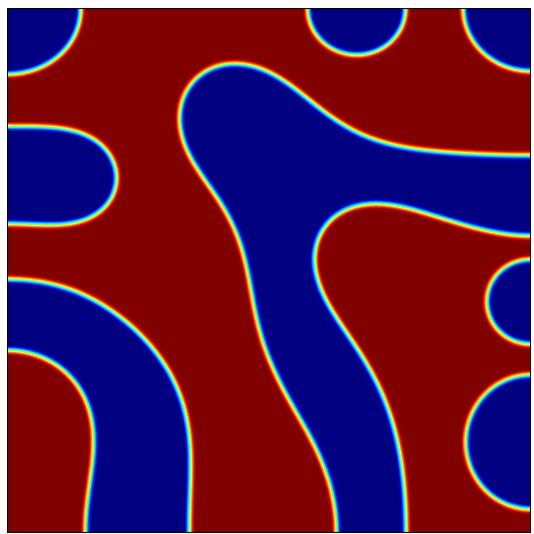}\\
\vspace{-7mm} \\
{\scriptsize  1 month} & {\scriptsize 1 day} & {\scriptsize 1 week} \\ 
\vspace{-10mm} \\
& & \\
\includegraphics[width=35mm]{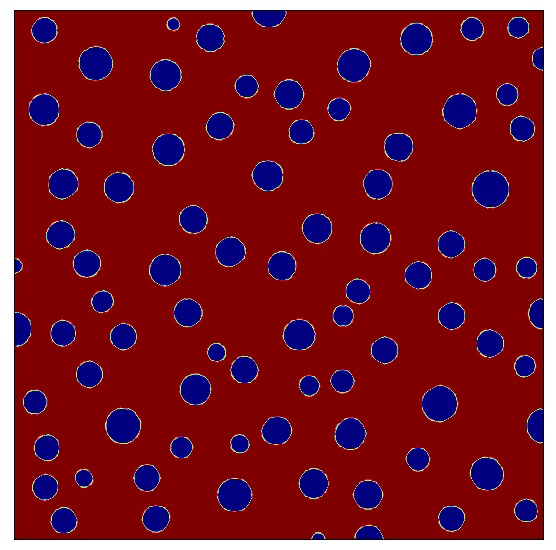}&
\includegraphics[width=35mm]{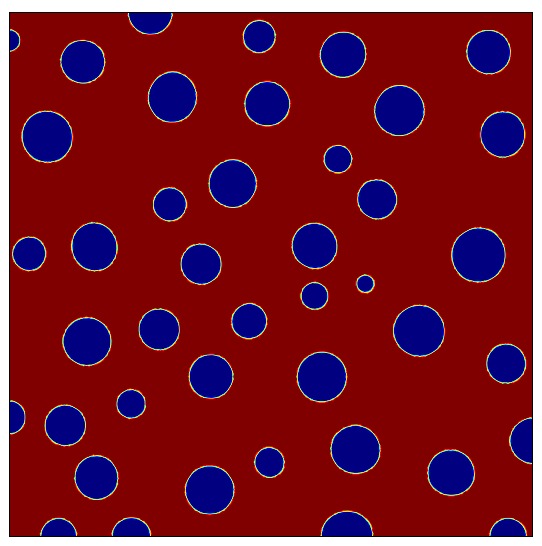}&
\includegraphics[width=35mm]{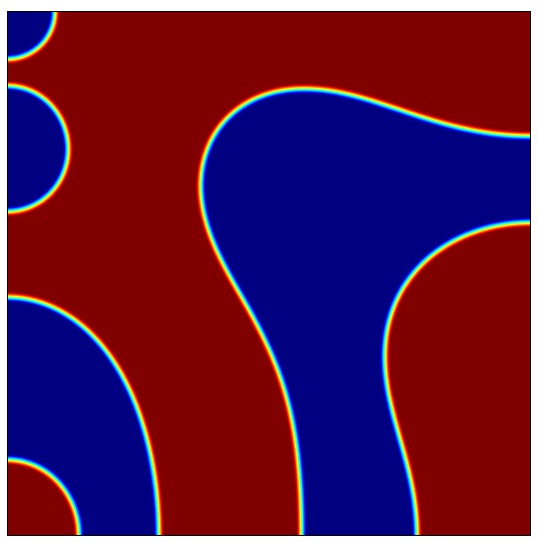}\\
\vspace{-7mm} \\
{\scriptsize  6 months} & {\scriptsize 1 week} & {\scriptsize  1 month}\\
\vspace{-10mm} \\
& & \\
\includegraphics[width=35mm]{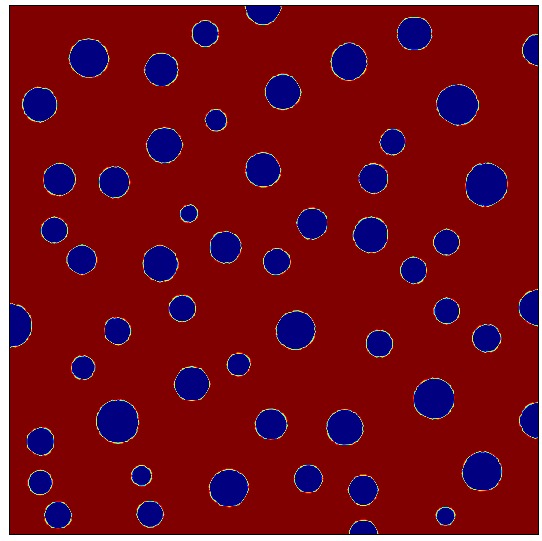}&
\includegraphics[width=35mm]{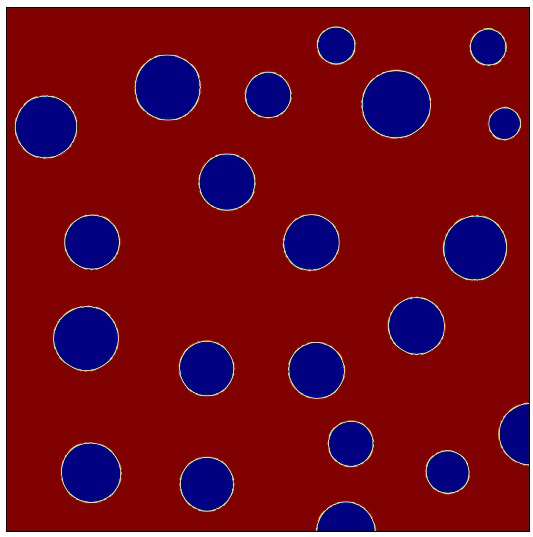}&
\includegraphics[width=35mm]{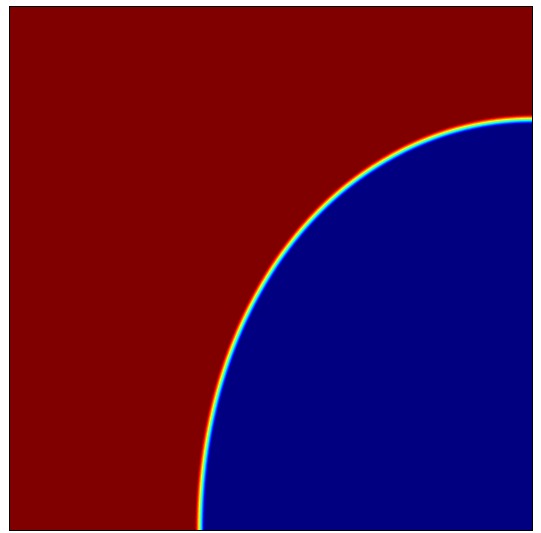}\\
\vspace{-7mm} \\
{\scriptsize 1 year} & {\scriptsize 1 month} & {\scriptsize  9 months}\\
\end{tabular}
\end{minipage}
\hspace{-20mm}
\begin{minipage}[b]{0.1\linewidth}
\begin{tabular}{c}
\\  \\ \\
\includegraphics[width=25mm]{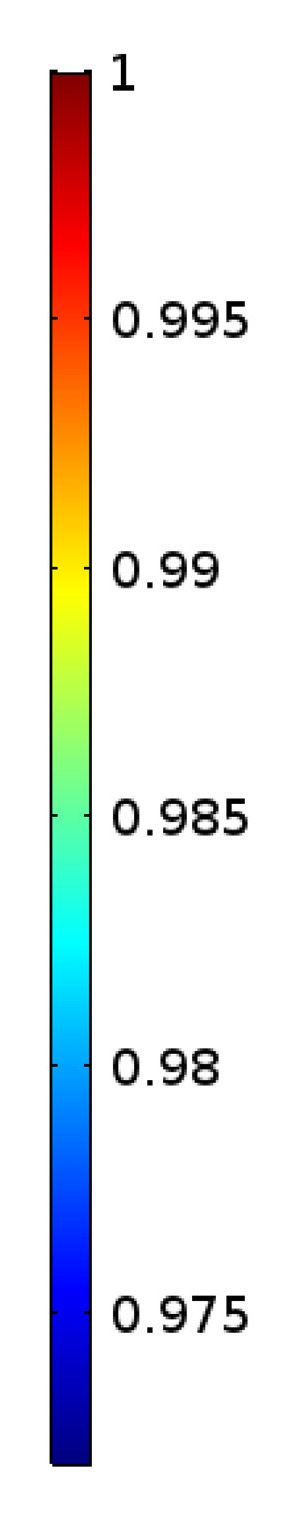} \\
\end{tabular}
\end{minipage}
\caption{See the caption for Figure \ref{fig:results80}. The weight 
              fraction of drug here is 60\%. }
\vspace{-10mm}
\label{fig:results60}
\end{figure}

\begin{figure}[p]
\centering 
\vspace{-27mm}
\begin{minipage}[b]{0.9\linewidth}
\begin{tabular}{ccc}
$T=60^{\circ}C$ & $T=90^{\circ}$ C & $T=110^{\circ} C$ \\
\includegraphics[width=35mm]{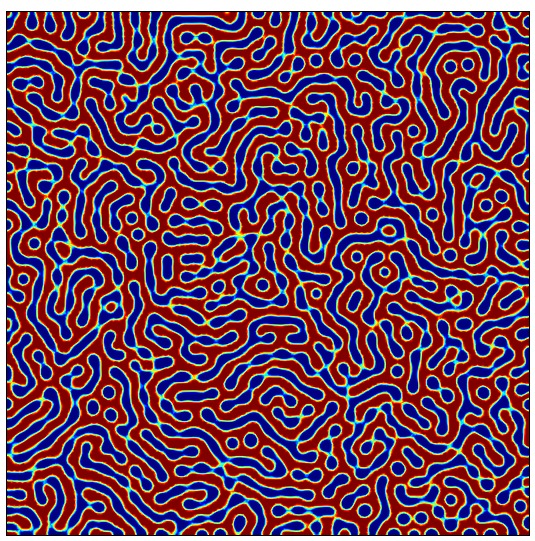} & 
\includegraphics[width=35mm]{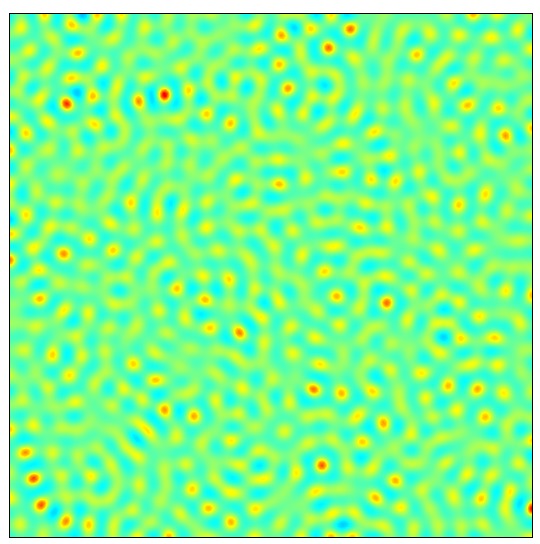} &
\includegraphics[width=35mm]{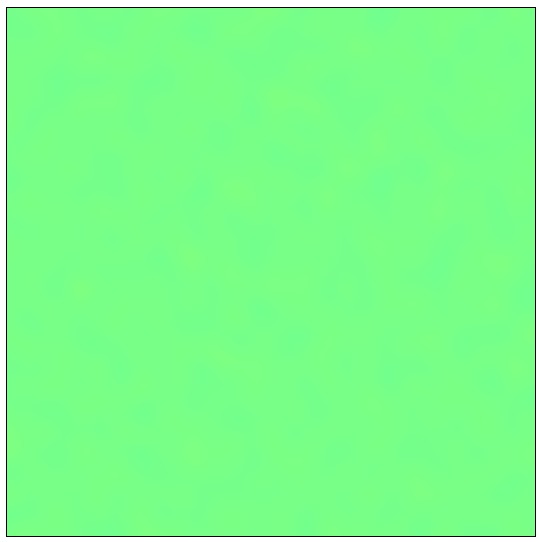} \\
\vspace{-7mm} \\
{\scriptsize  1 week} & {\scriptsize 3 hours} & {\scriptsize 3 hours}\\
\vspace{-10mm} \\
& & \\
\includegraphics[width=35mm]{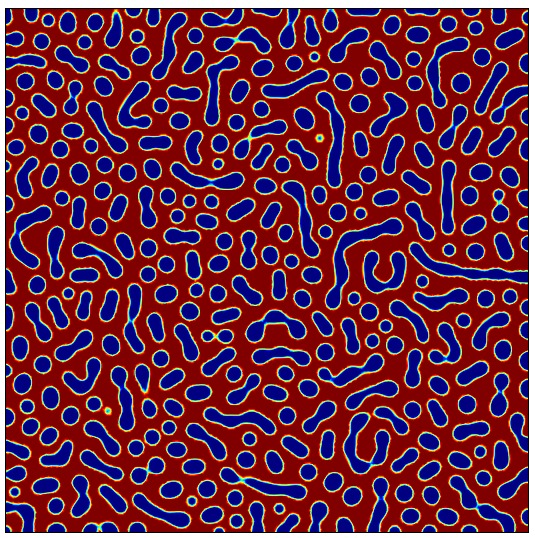}&
\includegraphics[width=35mm]{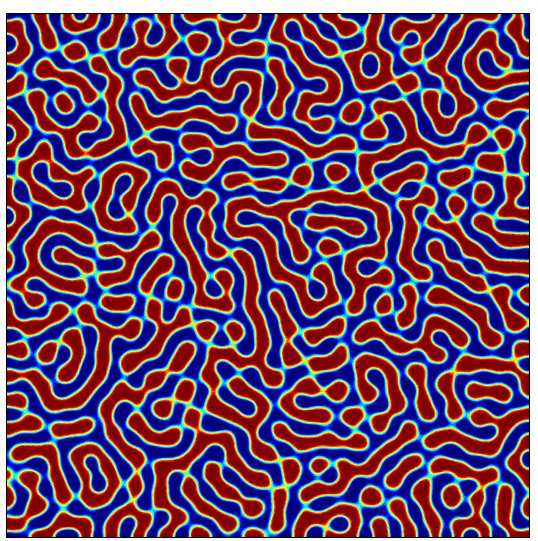}&
\includegraphics[width=35mm]{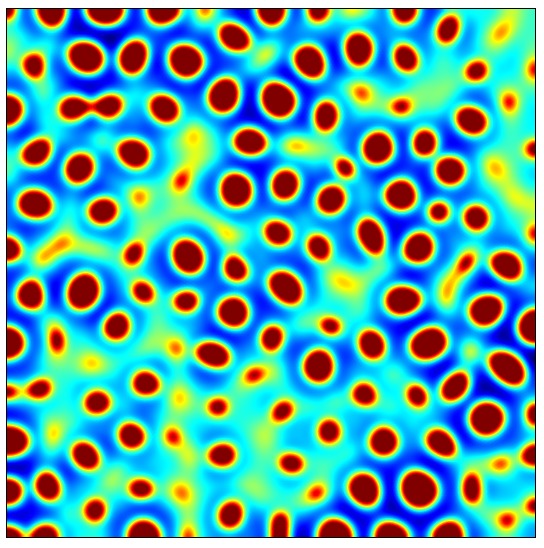}\\
\vspace{-7mm} \\
{\scriptsize 2 weeks} & {\scriptsize 6 hours} & {\scriptsize 6 hours}\\
\vspace{-10mm} \\
& & \\
\includegraphics[width=35mm]{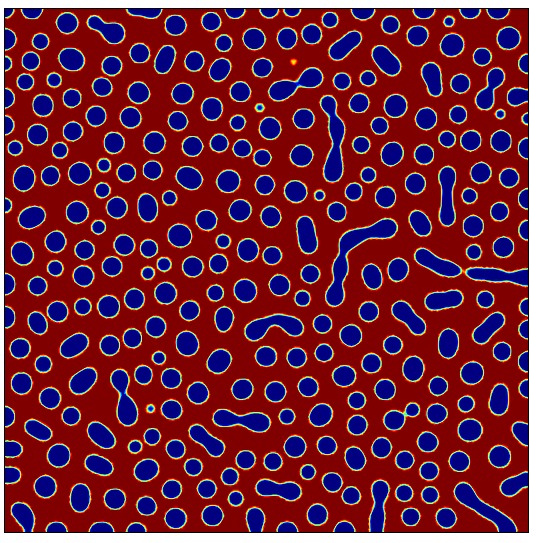}&
\includegraphics[width=35mm]{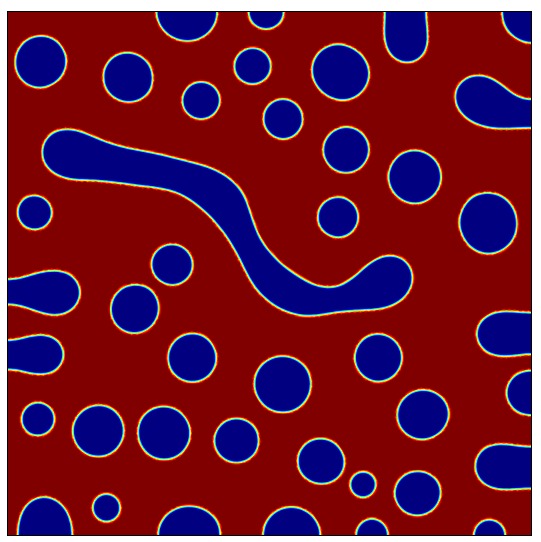}&
\includegraphics[width=35mm]{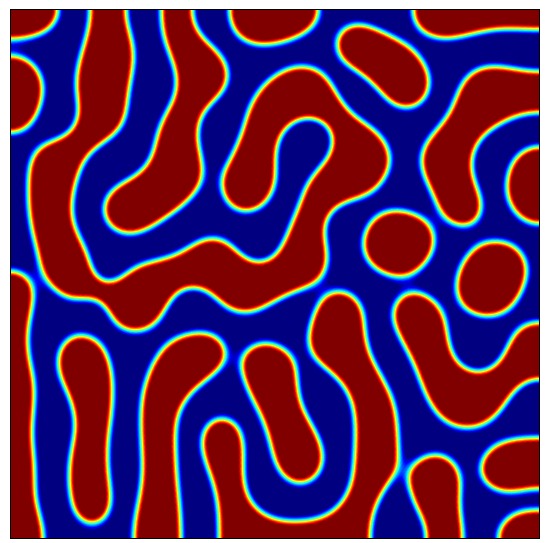}\\
\vspace{-7mm} \\
{\scriptsize 1 month} & {\scriptsize 1 week} & {\scriptsize 1 day} \\
\vspace{-10mm} \\
& & \\
\includegraphics[width=35mm]{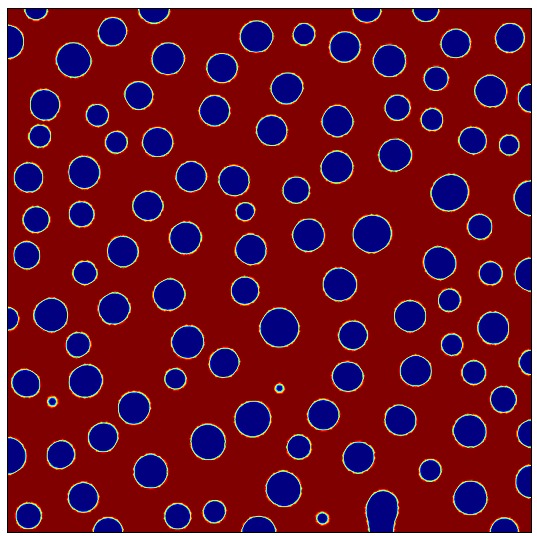}&
\includegraphics[width=35mm]{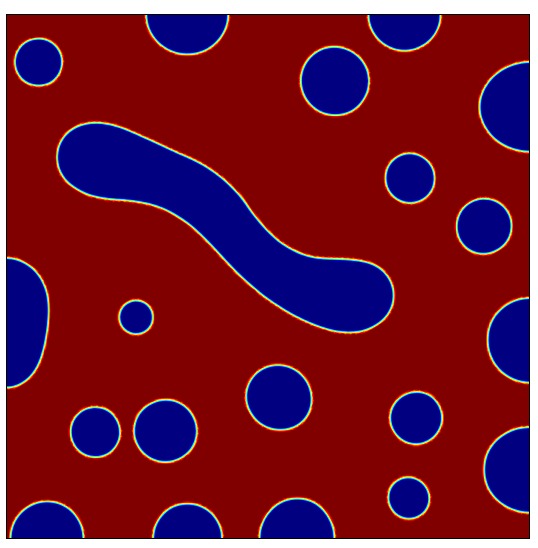}&
\includegraphics[width=35mm]{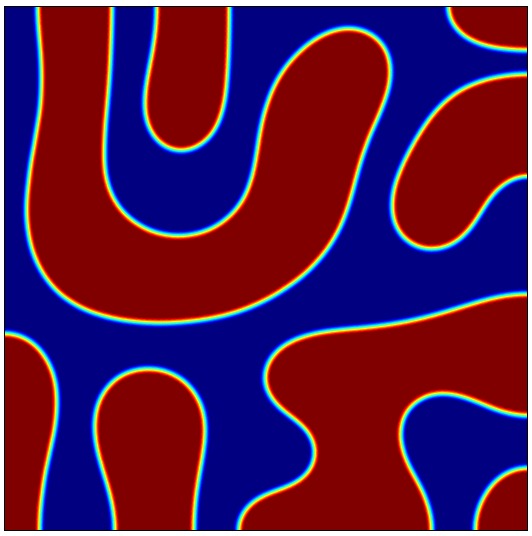}\\
\vspace{-7mm} \\
{\scriptsize 6 months} & {\scriptsize 1 month} & {\scriptsize 1 week}\\
\vspace{-10mm}\\ 
& & \\
\includegraphics[width=35mm]{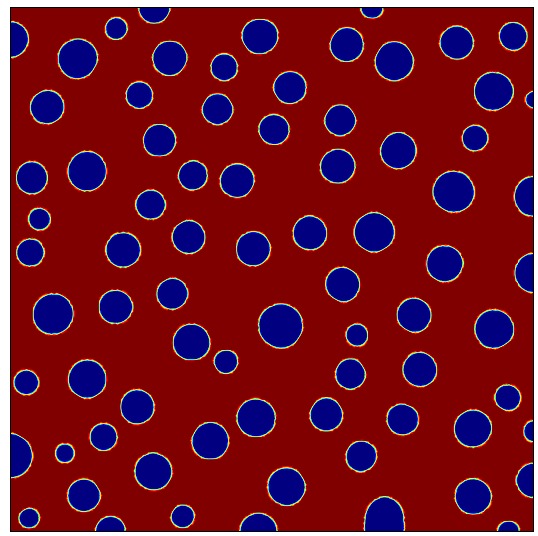}&
\includegraphics[width=35mm]{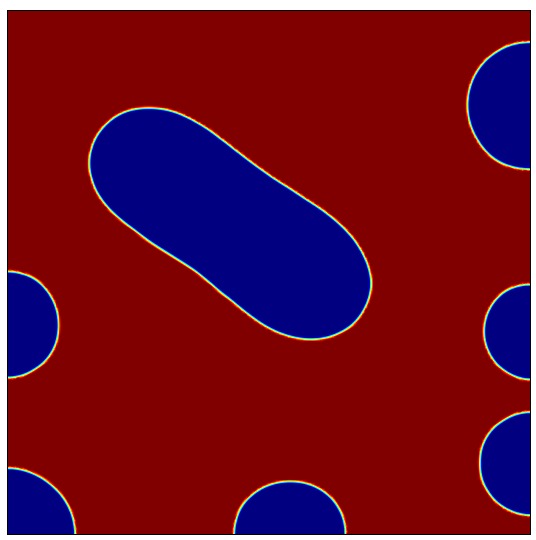}&
\includegraphics[width=35mm]{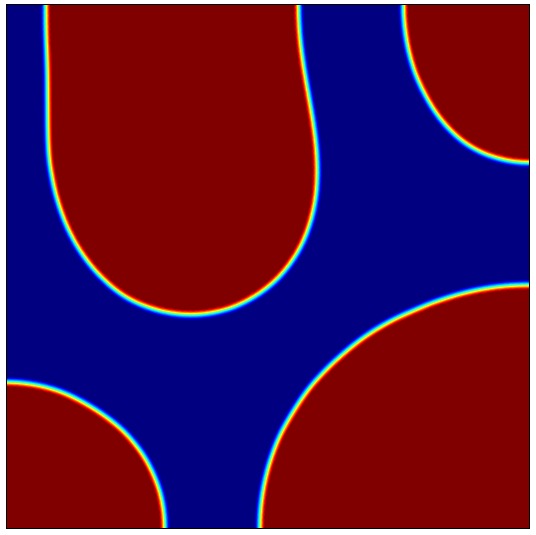}\\
\vspace{-7mm} \\
{\scriptsize 1 year} & {\scriptsize 6 months} & {\scriptsize 4 months}\\
\end{tabular}
\end{minipage}
\hspace{-20mm}
\begin{minipage}[b]{0.1\linewidth}
\begin{tabular}{c}
 \\ \\  \\
\includegraphics[width=25mm]{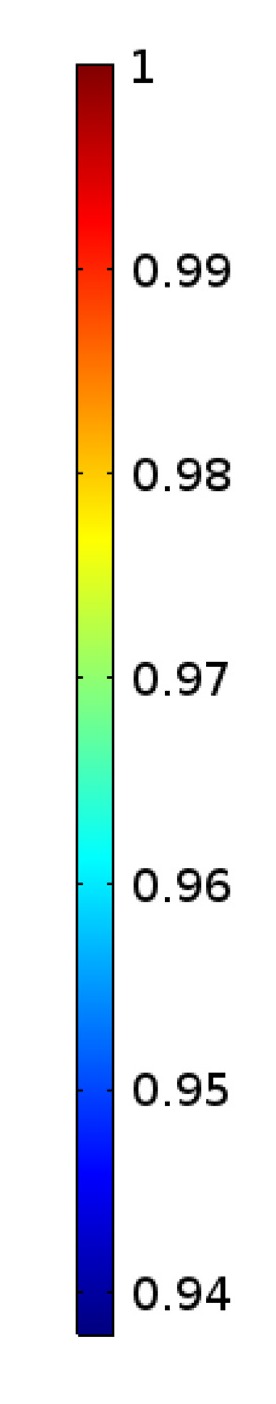} \\
\end{tabular}
\end{minipage}
\caption{See the caption for Figure \ref{fig:results80}. The weight 
             fraction of drug here is 40\%.}
\vspace{-10mm}
\label{fig:results40}
\end{figure}

\begin{figure}[p]
\centering 
\vspace{-27mm}
\begin{minipage}[b]{0.9\linewidth}
\begin{tabular}{ccc}
$T=60^{\circ}C$ & $T=75^{\circ}$ C & $T=90^{\circ} C$ \\
\includegraphics[width=35mm]{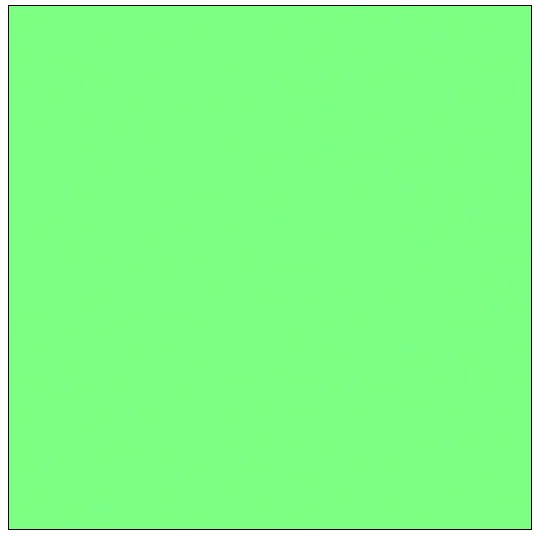} & 
\includegraphics[width=35mm]{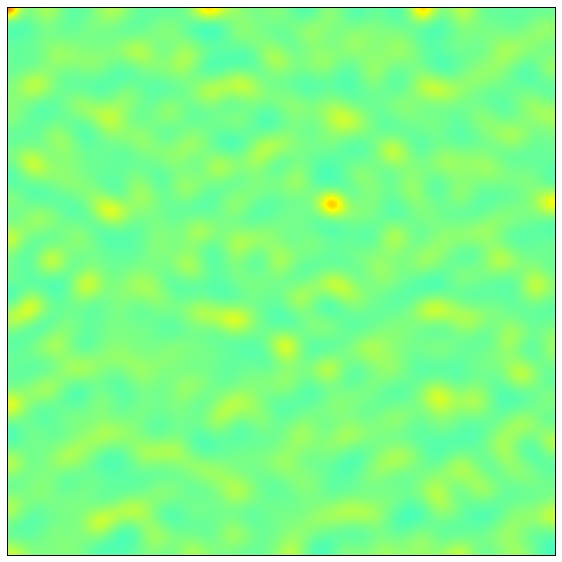} &
\includegraphics[width=35mm]{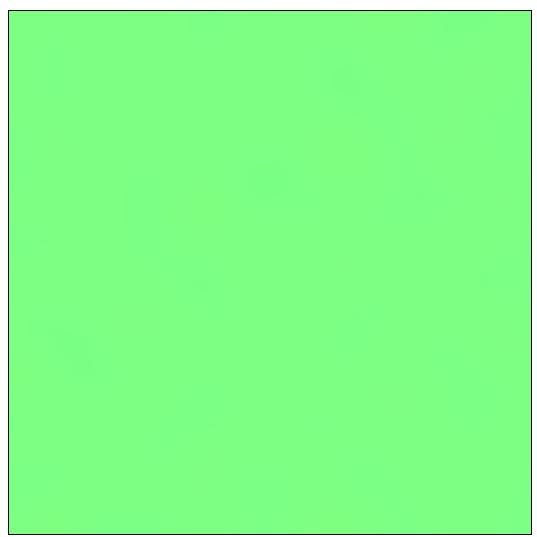} \\
\vspace{-7mm} \\
{\scriptsize  1 month} & {\scriptsize 2 weeks} & {\scriptsize 2 weeks}\\
\vspace{-10mm} \\
& & \\
\includegraphics[width=35mm]{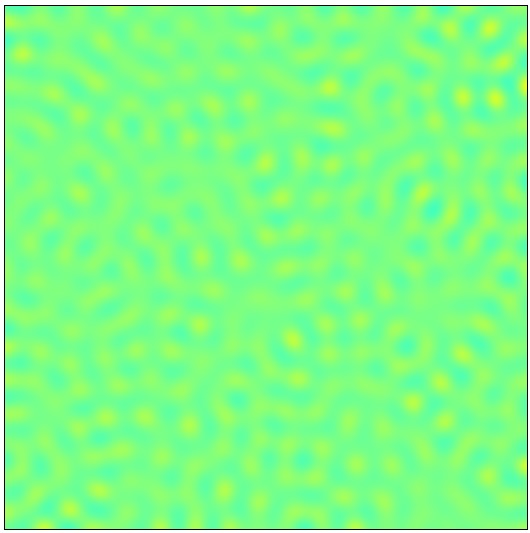}&
\includegraphics[width=35mm]{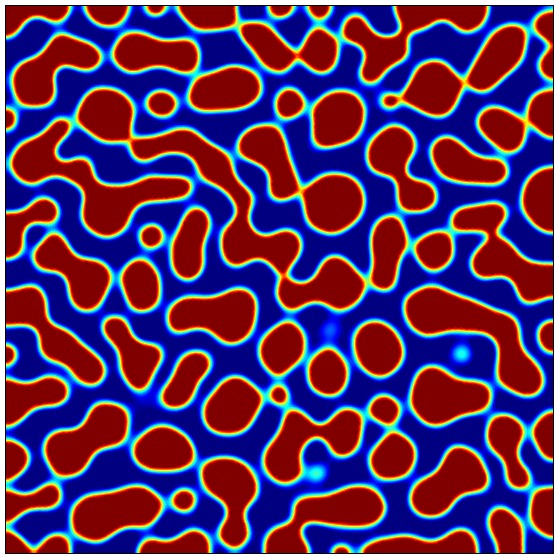}&
\includegraphics[width=35mm]{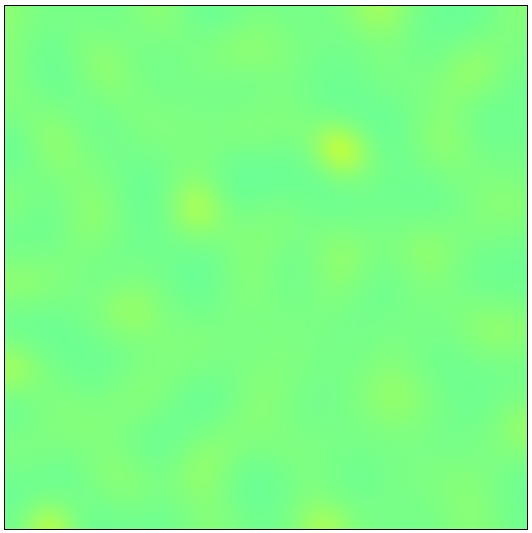}\\
\vspace{-7mm} \\
{\scriptsize 2 months} & {\scriptsize 23 days} & {\scriptsize 23 days}\\
\vspace{-10mm} \\
& & \\
\includegraphics[width=35mm]{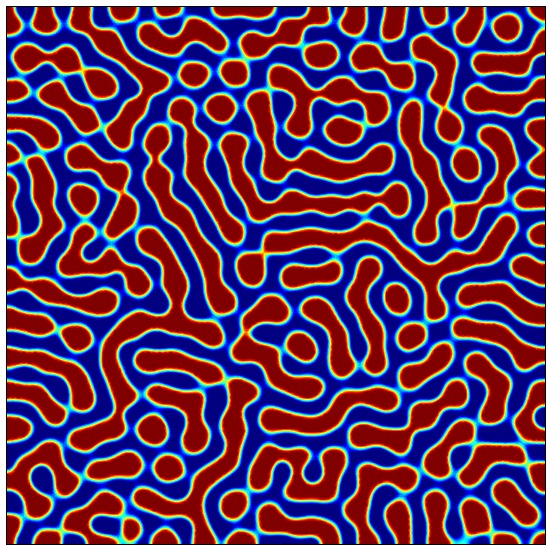}&
\includegraphics[width=35mm]{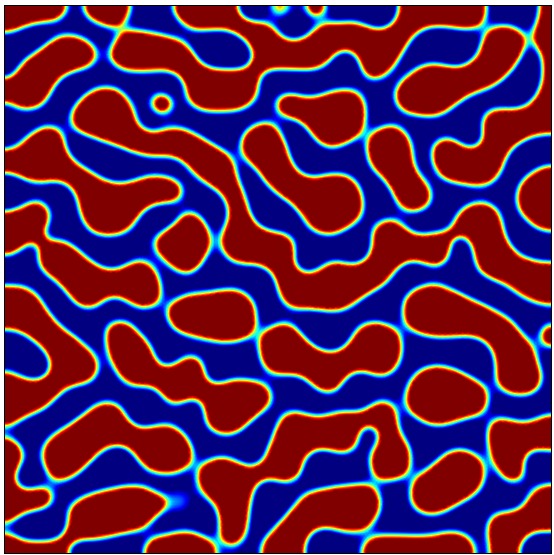}&
\includegraphics[width=35mm]{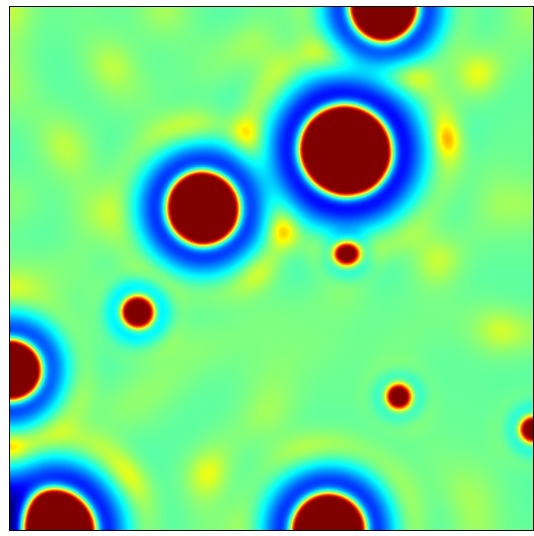}\\
\vspace{-7mm} \\
{\scriptsize 4 months} & {\scriptsize 1 month} & {\scriptsize 1 month} \\ 
\vspace{-10mm} \\
& & \\
\includegraphics[width=35mm]{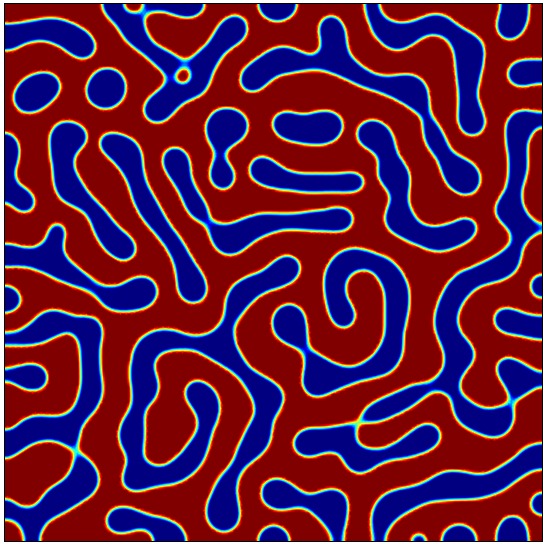}&
\includegraphics[width=35mm]{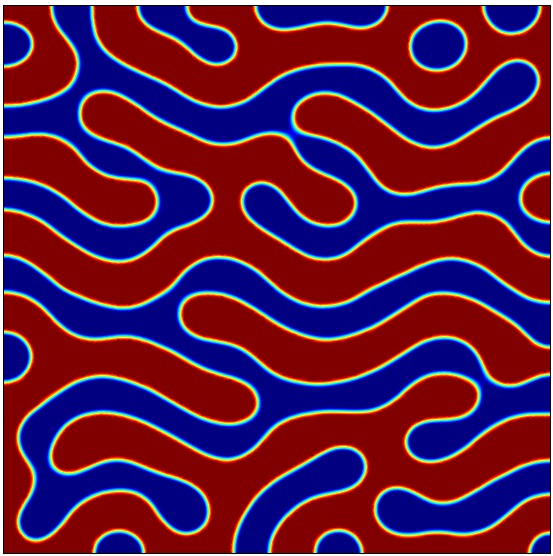}&
\includegraphics[width=35mm]{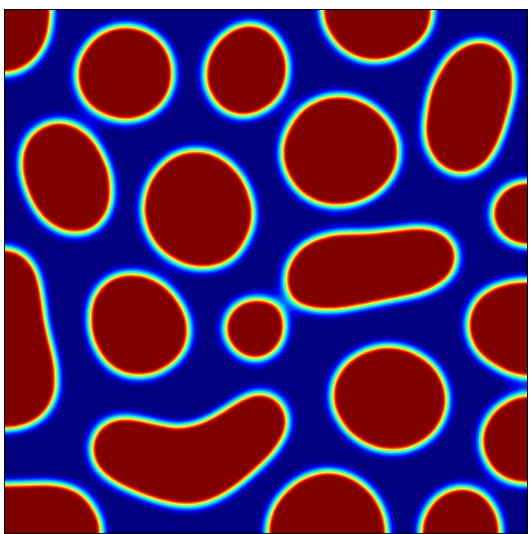}\\
\vspace{-7mm} \\
{\scriptsize 9 months} & {\scriptsize 2 months} & {\scriptsize 2 months}\\ 
\vspace{-10mm} \\
& & \\
\includegraphics[width=35mm]{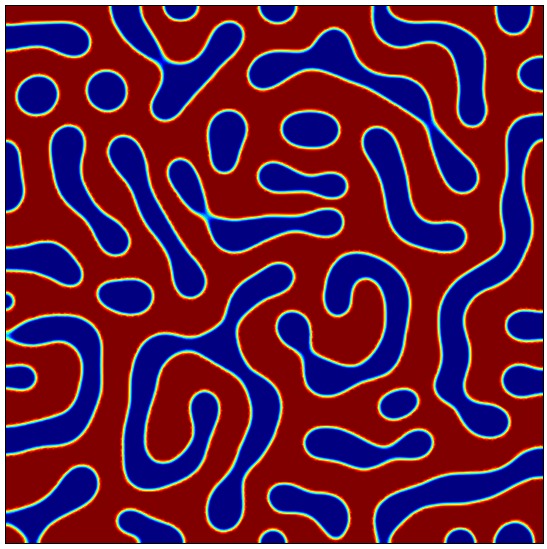}&
\includegraphics[width=35mm]{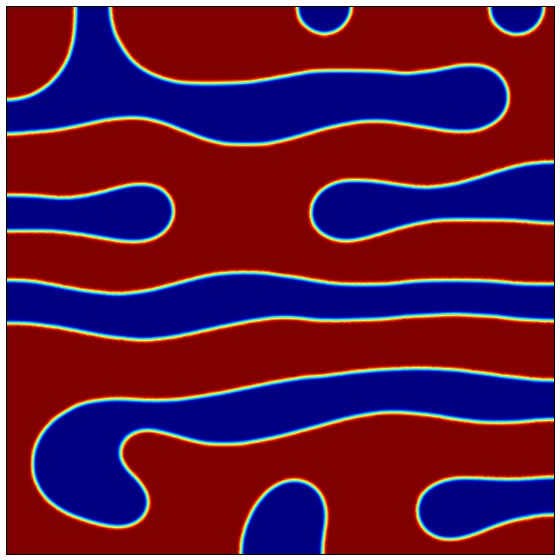}&
\includegraphics[width=35mm]{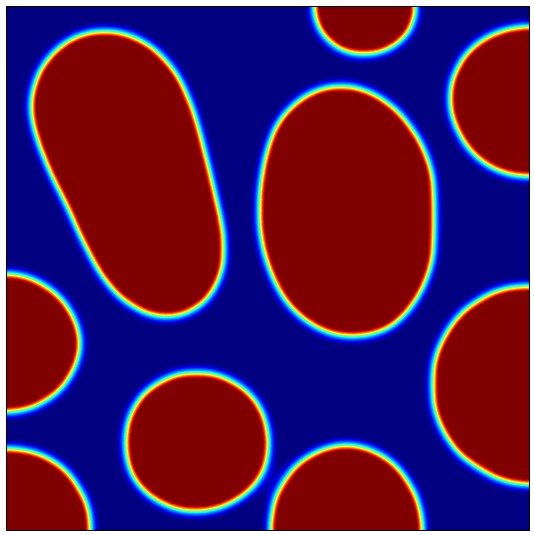}\\
\vspace{-7mm} \\
{\scriptsize 1 year} & {\scriptsize 9 months} & {\scriptsize 9 months}\\
\end{tabular}
\end{minipage}
\hspace{-20mm}
\begin{minipage}[b]{0.1\linewidth}
\begin{tabular}{c}
 \\ \\  \\
\includegraphics[width=25mm]{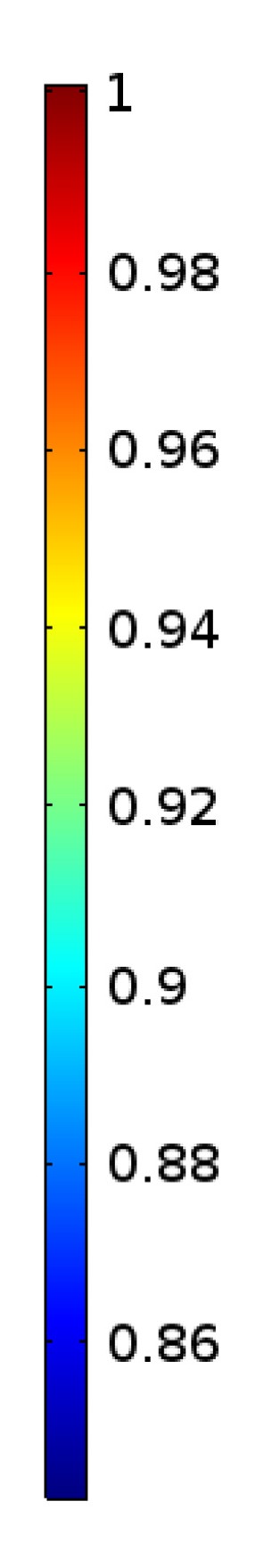} \\
\end{tabular}
\end{minipage}
\caption{See the caption for Figure \ref{fig:results80}. The weight 
              fraction of drug here is 20\%.}
\vspace{-10mm}
\label{fig:results20}
\end{figure}

Further numerical simulations are displayed in Figures  \ref{fig:results80},
\ref{fig:results60}, \ref{fig:results40}, \ref{fig:results20}, and these correspond to weight
fractions of drug of 80\%, 60\%, 40\% and 20\%, respectively. Recall that decreasing
weight fractions of drug correspond to increasing weight fractions of polymer since the
system is binary.  In a given figure, each column corresponds to a given temperature as 
labelled, and reading a column from top to bottom corresponds to increasing time for the 
dispersion for the given temperature. We have chosen here not to use the same times 
for the different temperatures since the rate at which a dispersion evolves depends on 
temperature. 

%%%%%%%%%%%%%%%%%%%%%%%%%%%%%%%%%%%%%%%%%%%

%%%%%%%%%%%%%%%%%%%%%%%%%%%%%%%%%%%%%%%%%%%

%%%%%%%%%%%%%%%%%%%%%%%%%%%%%%%%%%%%%%%%%%%

%%%%%%%%%%%%%%%%%%%%%%%%%%%%%%%%%%%%%%%%%%%

%%%%%%%%%%%%%%%%%%%%%%%%%%%%%%%%%%%%%%%%%%%

\begin{figure}
\centering 
\begin{minipage}[b]{0.9\linewidth}
\begin{tabular}{ccc}

\includegraphics[width=35mm]{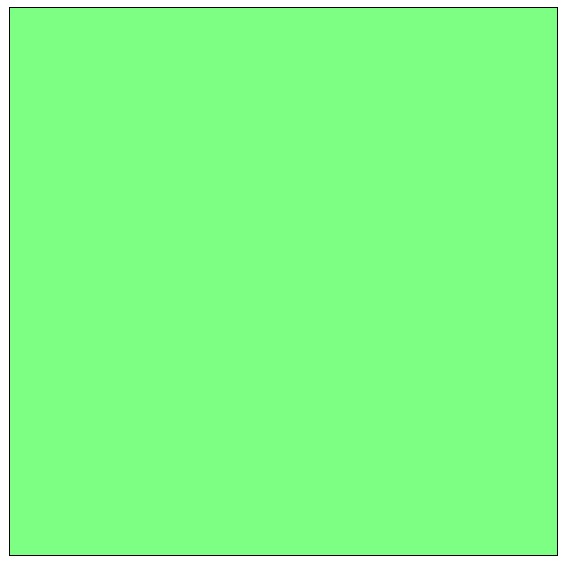} & 
\includegraphics[width=35mm]{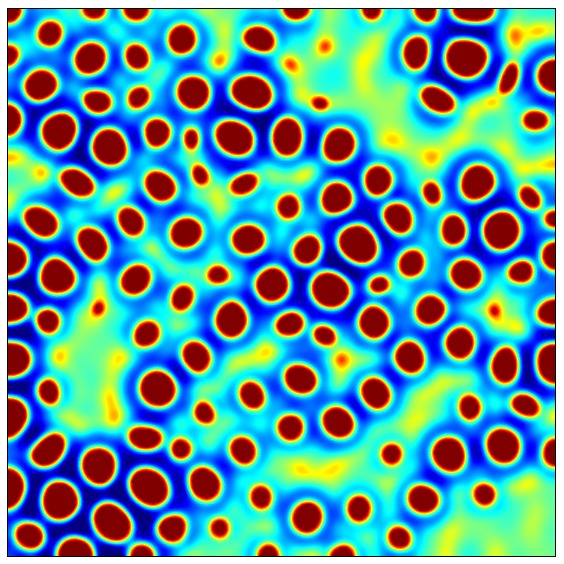} &
\includegraphics[width=35mm]{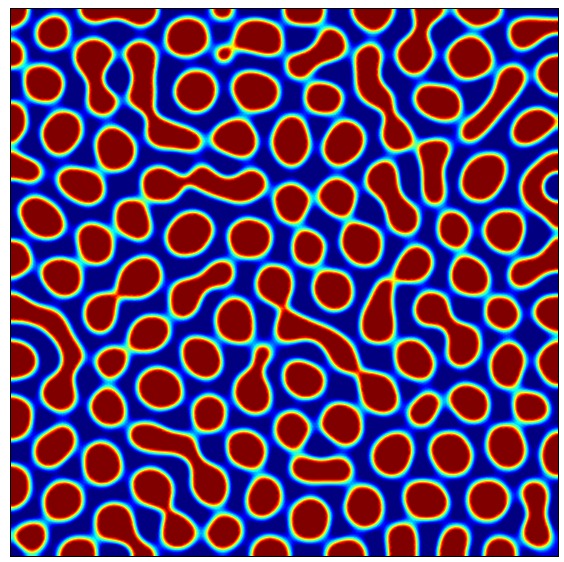} \\
\vspace{-7mm} \\
{\scriptsize  initial time} & {\scriptsize 16 days} & {\scriptsize 20 days}\\
\vspace{-10mm} \\
& & \\
\includegraphics[width=35mm]{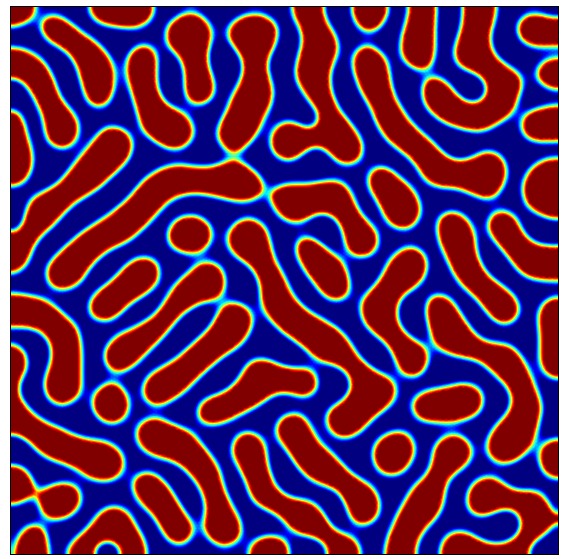}&
\includegraphics[width=35mm]{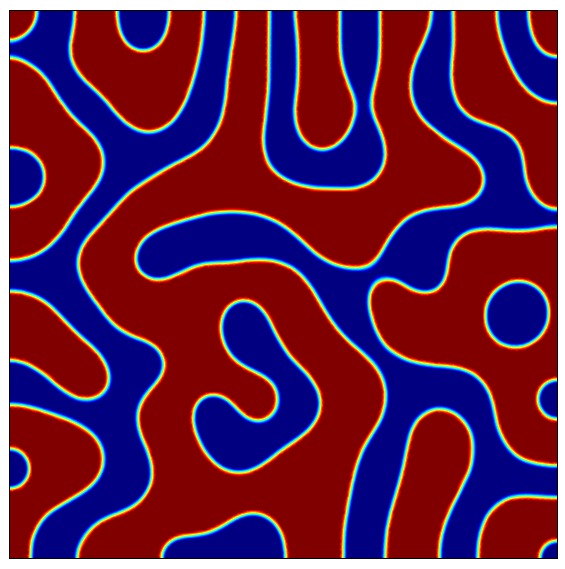}&
\includegraphics[width=35mm]{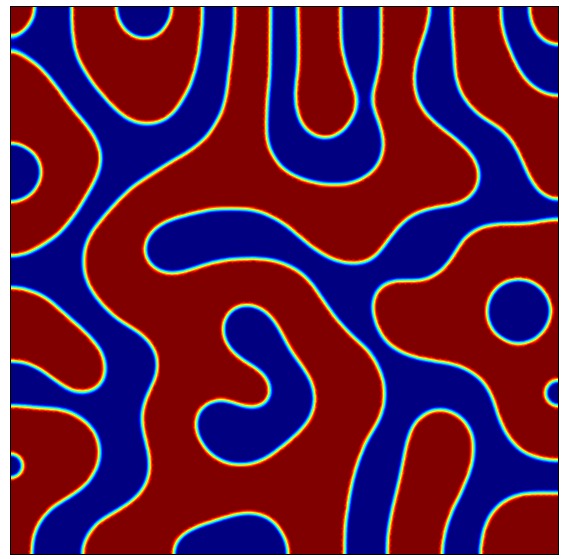}\\
\vspace{-7mm} \\
{\scriptsize 30 days} & {\scriptsize 100 days} & {\scriptsize 4 months}\\
\vspace{-10mm} \\
& & \\
\includegraphics[width=35mm]{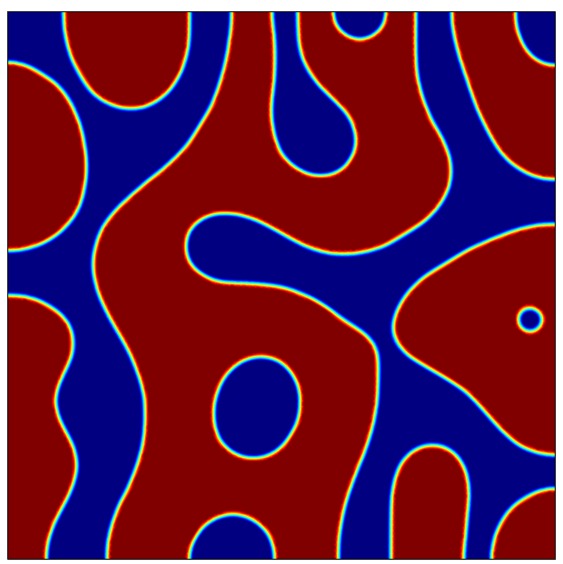}&
\includegraphics[width=35mm]{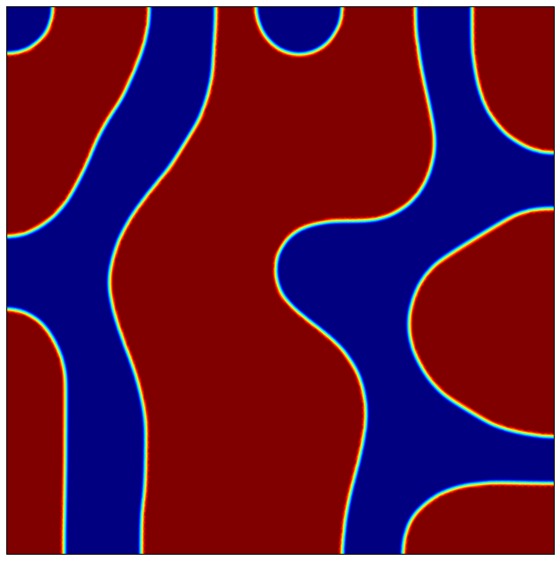}&
\includegraphics[width=35mm]{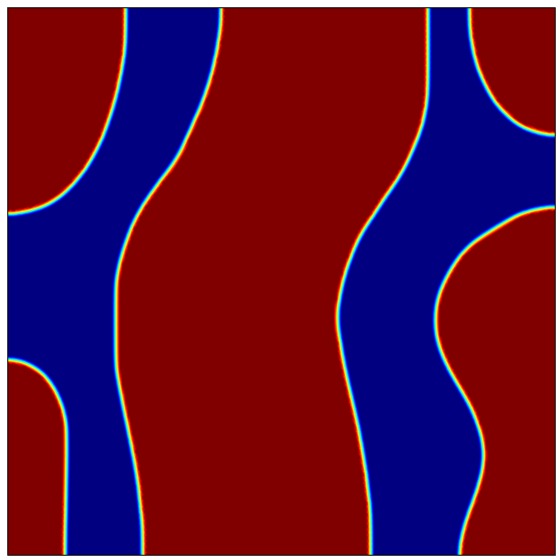}\\
\vspace{-7mm} \\
{\scriptsize 10 months} & {\scriptsize 36 months} & {\scriptsize 80 months} \\ 
\end{tabular}
\end{minipage}
\hspace{-20mm}
\begin{minipage}[b]{0.1\linewidth}
\begin{tabular}{c}
 \\ 
\includegraphics[width=18mm]{legend1921.jpg} \\
\end{tabular}
\end{minipage}
\caption{Numerical results illustrating a droplet-to-chain transition. In these
panels, the mass fraction of drug is 20\% and the temperature is $T=75^{\circ}C$.  
The panels should be read from left to right, starting at the top row. In the panel
for 16 days, we see the formation of drug droplets. The panels for 20 and 30 days
show the drug droplets in the process of chaining to form strings. Subesequent 
panels show the evolution of the drug strings.}
\vspace{-4mm}
\label{fig:dts2}
\end{figure}

We now highlight some notable features of these numerical simulations.
\begin{itemize}

\item {\em Two phases eventually emerge.} The numerical results show that the systems 
         eventually evolve into two distinct phases, characterized by deep blue 
         domains (polymer-rich) and deep red domains (drug-rich). 

\item {\em Ostwald ripening/coarsening.} Another notable feature in many of the numerical 
        illustrations is the formation of polymer droplets (blue discs) in the dispersion, followed 
        by a subsequent growth in their size; see, for example, the third column in 
        Figure \ref{fig:results80}.  This is a well-known and common phenomenon in multicomponent 
        solid systems, and is often referred to as Ostwald ripening or coarsening \cite{Ratke:2002}. 
        We also note the general trend that dispersions at higher temperature tend to be coarser.

\item{\em Phase inversion.} The system exhibits the phase inversion phenomenon \cite{Zhu:2016} 
        as the polymer content increases. To see this, consider the panels in Figure \ref{fig:results80}. 
        These correspond to the case where the polymer content is low (20\% by weight), and we see 
        the emergence of polymer droplets in drug-dominated domains. Compare these with the panels 
        in the third column of Figure \ref{fig:results20}. These correspond to the case where where 
        the polymer content is high (80\% by weight), and we see the emergence of drug droplets 
        in polymer-rich domains, the reverse of the low polymer content case.  

\item{\em Polymer strings and droplet-to-string transitions.} We note the formation of polymer
        strings in some of the panels; see the first and second columns of Figure \ref{fig:results20} 
        for examples. The central column in Figure \ref{fig:results20} is of particular interest since
        the behaviour exhibited here is an example of a droplet-to-string transition \cite{Migler:2001}. 
        In this droplet-to-string transition, drug droplets coalesce to form long drug-rich strings. In the 
        panel for 23 days, we observe that drug droplets are in the process of chaining \cite{Migler:2001}.
        Another droplet-to-string transition is shown in Figure \ref{fig:dts2}.  

\item{\em The formula (\ref{eq:diffusiontimescale}) for the timescale for phase separation.} The 
        detailed numerical results here enable us to test the utility of our simple formula 
        (\ref{eq:diffusiontimescale}) for the timescale for phase separation. Consider, for example, the 
        panel corresponding to 1 day in the third column of Figure \ref{fig:results80}. Here we see that 
        polymer droplets with characteristic lengthscale of $l\approx 0.3$ mm have formed. Our formula 
        (\ref{eq:diffusiontimescale}) predicts that such droplets should form over a timescale dictated by
       \[   \tau \approx 
            \frac{(0.3)^2 \mbox{mm}^{2}}{\vert \tilde{D}_{\mbox{\scriptsize eff}} (\phi_d=0.8006) \vert} 
                   \approx 11 \hspace{0.2cm} \mbox{hours}   \]
      which is consistent with the time $t=1$ day for the panel since 1 day $\approx 2\tau$. It should
      be emphasized that $\tau$ does not predict the time for the droplets to form, but rather estimates     
      the {\em timescale} over which such droplets form.      

\end{itemize}

\section{Conclusions}

Solid dispersions have been the subject of intensive research in recent years because of 
their potential to improve the solubility of drugs, and numerous excellent studies have been
published. However, detailed theoretical studies considering the non-equilibrium behaviour
of solid dispersions are lacking. Hence, in this study we have developed a general diffusion 
model for a dissolving solid dispersion. We then considered the particular case of a binary
system modelling a solid dispersion in storage, and developed a formula for the effective 
diffusion coefficient of the drug. We then specialized further to the case of a Flory-Huggins
statistical model. Within the context of this theory, we make the following predictions, some
of which should be testable experimentally: 

\begin{enumerate}

\item A solid dispersion can always be made stable by choosing a sufficiently low drug 
        loading; see Figure \ref{fig:Mobility} (a).  
        
\item For unstable regimes, the relationship between the local drug volume fraction 
        $\phi_{d}$ and the rate of phase separation is not obvious; see 
        Figure \ref{fig:Mobility} (a). There is in fact a well-defined value of $\phi_{d}$ 
        that corresponds to the most rapid rate of phase separation, with the rate decreasing 
        for values of $\phi_{d}$ either side of this value.

\item For unstable regimes, the rate of phase separation increases with increasing polymer 
        chain length $m$; see Figure \ref{fig:Mobility} (a).
        
\item Dispersions become more unstable with increasing value of the Flory-Huggins interaction
        parameter $\chi_{dp}$; see Figure \ref{fig:Mobility} (b).          

\item Binary drug/polymer systems are capable of exhibiting a rich set of dynamical behaviours. 
        In the numerical simulations performed in the current study, we observed the formation
        of polymer droplets and strings, the phase inversion phenomenon, Ostwald ripening, and 
        droplet-to-string transitions.         

\end{enumerate} 

The model can be evaluated empirically using microscopy by comparing 
the theoretical simulations with corresponding images seen in the microscope. 
Hot-stage polarized light microscopy is one notable possibility - see \cite{Liu:2018} 
for a discussion of relevant experimental techniques.

There is ample scope for extending the modelling work presented in the current study. One 
limitation of the binary model considered here is that it assumes that the polymer is 
perfectly dry. However, if the dispersions are stored in humid conditions, this is not a good 
assumption since even small amounts of moisture in the dispersion may significantly affect 
the mobility of the drug. Another avenue for extending the modelling work developed here 
is to use statistical models that capture more of the detail of the drug-polymer interaction
in the dispersion; see, for example, SAFT models \cite{Georgis:2010}. Viscoelastic 
effects may also play a significant role in the separation process since the polymer molecules 
are much larger than the drug molecules in a solid dispersion, giving rise to dynamic 
asymmetry between the components. Such models are significantly more complex than the model 
we have considered in the current study; see \cite{Tanaka:2000} for some discussion of such 
models. Another valid critique of the current modelling is that it is incapable of distinguishing 
between crystalline and amorphous drug.  Finally, the we have only considered the storage 
problem here, and have not addressed the dissolution behaviour at all. The dissolution of solid 
dispersions is at best partially understood, and there are many open issues that mathematical 
modelling may help resolve.  

It is noteworthy that the current study is the first (that we are aware of)   
that models in detail the spatiotemporal evolution of solid dispersions. Another novel 
feature of the current study is the development of an effective diffusion coefficient for 
the drug in the dispersion, the utility of which has been demonstrated in the results 
sections. An unusual feature of our modelling is that in equation (\ref{eq:Mici}) for the 
flux of species $i$, we have included the concentration $c_i$ of species $i$. This 
concentration term is frequently omitted in other studies, and a compositionally 
dependent mobility is assumed instead – see, for example, 
\cite{Zhu:2017} or \cite{Zhu:2016}.

\vspace{0.2cm}

\section*{Acknowledgments}
We thank the reviewers for their numerous helpful suggestions to improve the paper. 
The authors are grateful to S. Succi for fruitful discussions and acknowledge funding 
from the European Research Council under the European Unions Horizon 2020 Framework 
Programme (No. FP/2014-2020)/ERC Grant Agreement No. 739964 (COPMAT). M. Meere 
thanks NUI Galway for the award of a travel grant. 

\vspace{1.0cm} 

\noindent{\bf Appendix A. Linearization and stability analysis}

\vspace{0.3cm}

This appendix can be found in the supplementary material. 

%\bibliographystyle{elsarticle-num}

%\bibliography{refs}

\end{document}